\documentclass[twocolumn,english,aps,pra,superscriptaddress,showpacs,showkeys,longbibliography]{revtex4-1}
\usepackage[T1]{fontenc}
\usepackage[latin9]{inputenc}
\usepackage{color}
\usepackage{babel}
\usepackage{amssymb}
\usepackage{amsmath}
\usepackage{graphicx}
\usepackage[colorlinks=true,
    linkcolor=blue,
    citecolor=blue,
    urlcolor =blue]{hyperref}
\makeatletter
\usepackage{amsfonts}
\usepackage{babel}
\usepackage{braket}
\newcommand\minus{
  \setbox0=\hbox{-}
  \vcenter{
    \hrule width\wd0 height \the\fontdimen8\textfont3
  }%
}

\def\downket{\,\mid\downarrow\rangle}
\def\upbra{\langle\uparrow\mid}

\definecolor{mygreen}{rgb}{0,0.5,0}
\definecolor{myblue}{rgb}{0,0,0.75}
\definecolor{mymagenta}{cmyk}{0,1,0,0.12}

\usepackage{umoline}

\makeatother

\begin{document}

\title{Non-Markovian Dynamics in Chiral Quantum Networks with Spins and Photons}

\author{Tom\'as Ramos}
\thanks{These two authors contributed equally}
\affiliation{Institute for Quantum Optics and Quantum Information of the Austrian
Academy of Sciences, A-6020 Innsbruck, Austria}
\affiliation{Institute for Theoretical Physics, University of Innsbruck, A-6020
Innsbruck, Austria}

\author{Beno\^it Vermersch}
\thanks{These two authors contributed equally}
\affiliation{Institute for Quantum Optics and Quantum Information of the Austrian
Academy of Sciences, A-6020 Innsbruck, Austria}
\affiliation{Institute for Theoretical Physics, University of Innsbruck, A-6020
Innsbruck, Austria}

\author{Philipp Hauke}
\affiliation{Institute for Quantum Optics and Quantum Information of the Austrian
Academy of Sciences, A-6020 Innsbruck, Austria}
\affiliation{Institute for Theoretical Physics, University of Innsbruck, A-6020
Innsbruck, Austria}

\author{Hannes Pichler}
\affiliation{Institute for Quantum Optics and Quantum Information of the Austrian
Academy of Sciences, A-6020 Innsbruck, Austria}
\affiliation{Institute for Theoretical Physics, University of Innsbruck, A-6020
Innsbruck, Austria}

\author{Peter Zoller}
\affiliation{Institute for Quantum Optics and Quantum Information of the Austrian
Academy of Sciences, A-6020 Innsbruck, Austria}
\affiliation{Institute for Theoretical Physics, University of Innsbruck, A-6020
Innsbruck, Austria}

\begin{abstract} 
We study the dynamics of chiral quantum networks consisting of nodes coupled by unidirectional or asymmetric bidirectional quantum channels. In contrast to familiar photonic networks where driven two-level atoms exchange photons via 1D photonic nanostructures, we propose and study a setup where interactions between the atoms are mediated by spin excitations (magnons) in 1D $XX$ spin chains representing spin waveguides. While Markovian quantum network theory eliminates quantum channels as structureless reservoirs in a Born-Markov approximation to obtain a master equation for the nodes, we are interested in non-Markovian dynamics. This arises from the nonlinear character of the dispersion with band-edge effects, and from finite spin propagation velocities leading to time delays in interactions. To account for the non-Markovian dynamics we treat the quantum degrees of freedom of the nodes and connecting channels as a composite spin system with the surrounding of the quantum network as a Markovian bath, allowing for an efficient solution with time-dependent density matrix renormalization group techniques. We illustrate our approach showing non-Markovian effects in the driven-dissipative formation of quantum dimers, and we present examples for quantum information protocols involving quantum state transfer with engineered elements as basic building blocks of quantum spintronic circuits.
\end{abstract}
\maketitle

\section{Introduction} 

A quantum network consists of a set of nodes exchanging quantum information via connecting quantum channels. The prototypical examples are networks of small scale quantum computers as a local area network, to scale up quantum information processors; or wide-area networks, where the goal is long distance quantum communication in a quantum internet, involving transmission of quantum states with high fidelity ~\cite{SbKimble2008,SbRitter2012,SbNickerson:2014ci,SbHucul2014}. To implement long-distance communication, optical photons as {\em flying qubits} propagating in optical fibers, or free space, are the unchallenged physical realizations of quantum information carriers \cite{SbNorthup:2014gv}. In contrast, to implement a local `on-chip'  network~\cite{SbSchoelkopf2008}, there are several appealing options for {\em flying qubits} in both atomic and solid state settings, and corresponding interfaces to {\em stationary qubits} \footnote{An alternative mechanism for transfer of quantum states is a `shuttle', where qubits stored in atoms, ions, or quantum dots are physically transported in a quantum network \cite{SbKielpinski:2002di,SbTaylor:2005ip}.}. This includes optical photons combined with the possibilities offered by engineered nanostructures~\cite{SbTiecke2014,SbGoban:2015dr,SbLodahl:2015fy,SbVetsch:2010ev,SbSushkov:2014dz}, microwave photons in a superconducting strip-line~\cite{SbAstafiev2010,SbHoi2011,SbVanLoo2013,SbWang:2016ty}, and phonon waveguides~\cite{SbFang:2015vr,SbRiedinger:2016cl}, with a low temperature environment as a prerequisite. Another intriguing possibility is to use engineered spin chains as waveguides for magnons (spin excitations), which can carry quantum information or mediate interactions between the nodes of the quantum network~\cite{SbLoss1998,SbBose2007,SbCoish2008,Sbnikolopoulos2013quantum}.  

\begin{figure}[t]
\includegraphics[width=0.5\textwidth]{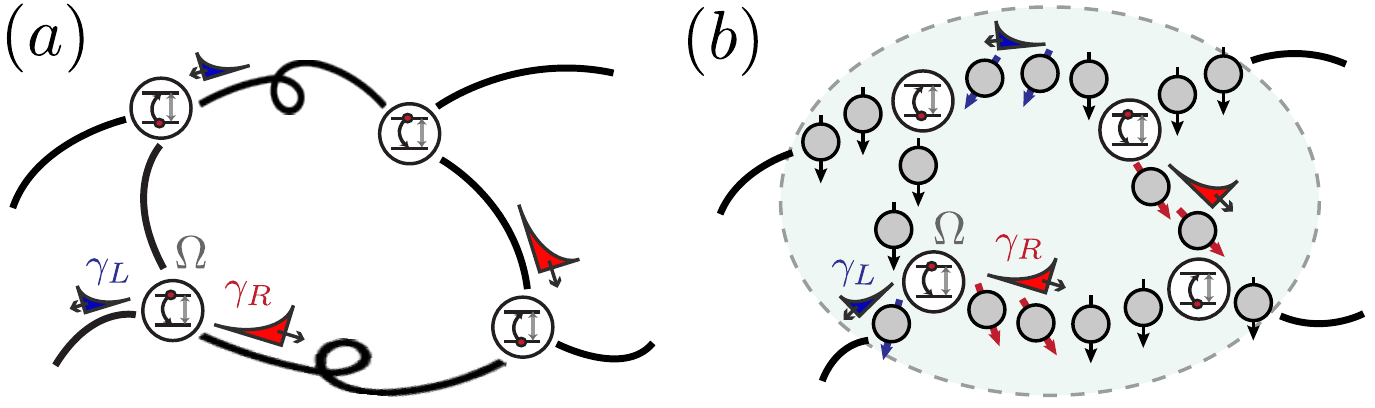}
\caption{(a) Photonic quantum network of two-level systems coupled to waveguides (optical fibers). (b) Quantum network with a spin chain representing a waveguide for spin excitations. To provide a systematic treatment of the quantum many-body dynamics beyond the Born-Markov approximation, we employ tDMRG techniques to treat the full dynamics of the two-level systems (nodes) and spin chains (quantum channels) within the region inside the dashed lines, thus defining an extended {\em Markovian cut}.}
\label{Sbfig:vision}
\end{figure}

Figure~\ref{Sbfig:vision}(a) illustrates a simple example of a photonic quantum network, where two-level atoms represent the nodes, which are connected by fibers allowing exchange of photons between the atoms. The counterpart to this setup, but now with a spin chain as quantum channel, is outlined in Fig.~\ref{Sbfig:vision}(b), where magnons as spin excitations in the spin waveguide take the role of {\em hard core} photons. The quantum networks of Fig.~\ref{Sbfig:vision} represent {\em open} quantum systems, where the photonic and spin channels also provide input and output ports for the quantum circuits.

In this work, we study the dynamics of such quantum networks from the perspective of an open many-body quantum system. An effective dynamics of the nodes alone is obtained by integrating out the degrees of freedom associated with the quantum channels, resulting in a reduced system dynamics that is in general {\em non-Markovian}~\cite{SbRivas2014,SbBreuer2015,SbdeVega:2015wi}. In a quantum-optical context, however, a Born--Markov approximation to trace out the photonic quantum channel as a structureless reservoir is often an excellent approximation, based on weak-coupling perturbation theory and the neglect of {\em time delays} in interactions (retardation)~\cite{Sbgardiner2014,Sbgardiner2015,SbLehmberg:1970jj}. This approximation leads to a {\em Markovian} theory of quantum networks, and an effective description of the reduced dynamics of the nodes in terms of a master equation (ME). However, for structured reservoirs such as a photonic bandgap material \cite{SbGoban:2015dr,SbGonzalezTudela:2000cy,SbDouglas2015}, or for a spin-chain implementation as in Fig.~\ref{Sbfig:vision}(b), the Bloch-band character with {\em band-edge effects}, and the {\em finite propagation speed} of spin excitations can invalidate a weak-coupling Markovian theory~\cite{SbJohn1990,SbKofman1994,SbNavarrete-Benlloch2011,SbDouglas2015,SbCalajo2015,SbChen2016,SbShi:2015uj,SbGonzalezTudela:2000cy,SbMiloni1974}. 

We wish to develop our theory for {\em chiral} quantum networks, and for chiral quantum networks with spin waveguides in particular. By chirality we mean a left--right asymmetry, or directionality in the emission of excitations from the quantum nodes into the quantum channels. In photonic quantum networks, chirality appears as a natural and generic manifestation of spin-orbit coupling of light \cite{SbBliokh:2015bw}, as recently demonstrated in seminal experiments with atoms or quantum dots coupled to photonic nanostructures~\cite{SbPetersen2014,SbMitsch2014,SbSollner2015,SbYoung2015,SbColes:2016dw,SbNeugebauer:2014iy,SbRodriguezFortuno:2013ew,SbLin:2013do,SbGorodetski:2012cg}. We will show below that chiral couplings can also be engineered for quantum networks with spin waveguides, as an essential ingredient in spin network design. Chirality of quantum channels in dynamics of quantum networks leads to several qualitatively new phenomena, such as the formation of quantum dimers as pure `dark' steady states of the driven-dissipative dynamics \cite{SbStannigel:2012jk,SbRamos2014,SbPichler2015}. 
In addition, chirality provides interesting opportunities and applications in a quantum information context, allowing for the design of `non-reciprocal' circuits \cite{SbHabraken:2012in,SbMetelmann:2015gb} in a form of {\em chiral quantum spintronics}. 
We emphasize that quantum dimer formation and quantum state transfer between nodes \cite{SbCirac1997,SbStannigel2011} have been discussed under {\em Markovian} assumptions so far. 

Designing and describing a {\em chiral quantum network with spin waveguides} encounters several challenges and questions, and in particular also differences with respect to chiral photonic networks. The present work serves to analyze the physics of chiral spin networks in detail. We discuss how to engineer a chiral coupling of the nodes into the spin waveguide, which is accomplished by adding synthetic gauge fields to the interactions between node and waveguide. 
The main focus of the work is then on analyzing the dynamics of chiral spin networks, with a particular emphasis on non-Markovian effects and witnesses of non-Markovianity, especially in the driven-dissipative formation of quantum dimers. Moreover, we present examples for potential applications of nonlinear quantum spin circuits in quantum-information protocols, as a quantum form of spintronics~\cite{SbLenk2011,SbChumak2015,SbKarenowska2015}.
In an accompanying paper \cite{SbImplementation}, we give details for various physical implementations based on dipolar interactions in arrays of Rydberg atoms or polar molecules~\cite{SbYao2012,SbSyzranov2014,SbPeter2015}, and comment on possible solid-state realizations, also in light of potential advantages of spin setups. A study of spin networks is also timely in view of the recent progress and new possibilities in engineering spin chains and spin quantum channels with atomic and molecular ensembles, and with solid-state impurities. 

To address the question of non-Markovian dynamics, we treat the chiral quantum network of nodes coupled to the connecting spin chains as one large quantum many-body system, keeping the entanglement between the nodes and the (possibly many) excitations propagating in the quantum channel, while treating the surrounding of the quantum network as a Markovian bath for the network dynamics [cf.~Fig.~\ref{Sbfig:vision}(b)]. This treatment is in the spirit of representing non-Markovian quantum stochastic processes as a {\em projection} of a quantum Markov process for an extended, but still finite number of degrees of freedom \cite{SbBreuer:2004fc,SbWoods:2014kn,SbHughes2009}. To solve the {\em extended master equation} for the quantum network as a driven-dissipative quantum many-body system, we exploit the fact that our formalism embodies the network as a quasi-one-dimensional lattice system. This allows us to employ efficient tools such as the time-dependent density matrix renormalization group (tDMRG) techniques~\cite{SbSchollwock2011}, adapted to open quantum systems~\cite{SbVerstraete:2004hi,SbDaley2014,SbWerner:2014wq,SbCui:2015im,SbMascarenhas:2015cc}. 
We remark that this approach also provides full access to the dynamics of excitations propagating in the quantum channels on the level of a quantum many-body wavefunction.

While the present work focuses on quantum networks with spin chains, we emphasize that this treatment of strong-coupling non-Markovian effects also carries over to photonic circuits with bandgap materials \cite{SbGoban:2015dr,SbGonzalezTudela:2000cy,SbDouglas2015}, or coupled cavity arrays \cite{SbSanchezBurillo:2016vg,SbLombardo:2014ju}. In this sense, modern quantum many-body techniques based on matrix product states become a significant tool to solve for dynamics of spin and photonic quantum networks, beyond the paradigm of the Born--Markov approximation in quantum optics. 

The paper is organized as follows. In Sec.~\ref{Sbsec:CQN} we define our model of a chiral quantum network. We explain the chiral coupling of nodes to spin waveguides, discuss limiting cases where the model is equivalent to a photonic network, and introduce the extended Markovian cut allowing us to describe the non-Markovian dynamics. This is the core part of our theoretical description. In Sec.~\ref{Sbsec:NonMark}, we illustrate and analyze various non-Markovian effects, due to a finite Bloch band, time delays, strong coupling, and spin hard-core effects, in particular in the formation of quantum dimers in a driven-dissipative system. Finally, Sec.~\ref{Sbsec:spintronics} presents examples on how nonlinear quantum circuits with spins, as a form of quantum spintronics, may be applied in quantum information protocols. 

\section{Chiral Quantum Network Model with Spin Waveguides}\label{Sbsec:CQN}

In this section we define our model of a quantum network with chiral coupling of nodes to spin waveguides as quantum channels. Below we will first illustrate the mechanism behind the chiral coupling [cf.~Sec.~\ref{Sbsec:model}]. To establish the relation with previous work, we then show that our model reduces to the chiral photonic network model~\cite{SbRamos2014,SbPichler2015} in the limit of weak coupling and low spin excitation density [cf.~Sec.~\ref{Sbchirality_infinite}]. Furthermore, applying the Born-Markov approximation allows us to eliminate the spin waveguide as a structureless reservoir, and thus to derive a master equation for the nodes (two-level systems) defining a {\em Markovian chiral network} [cf.~Sec.~\ref{Sbsec:markov}]. Finally, in preparation for our discussion of the strong-coupling limit and non-Markovian dynamics, we derive in Sec.~\ref{Sbsub:beyondMarkExt} an {\em extended master equation}. This is obtained by keeping not only the nodes but also the connecting spin waveguides as part of an `extended system' dynamics on the level of a many-body wavefunction. Thus, we effectively move the {\em Markovian cut} in the quantum channel all the way out to the input and output ports of the network, as indicated by the dashed lines in Fig.~\ref{Sbfig:vision}(b). 

\begin{figure}[b]
\includegraphics[width=0.5\textwidth]{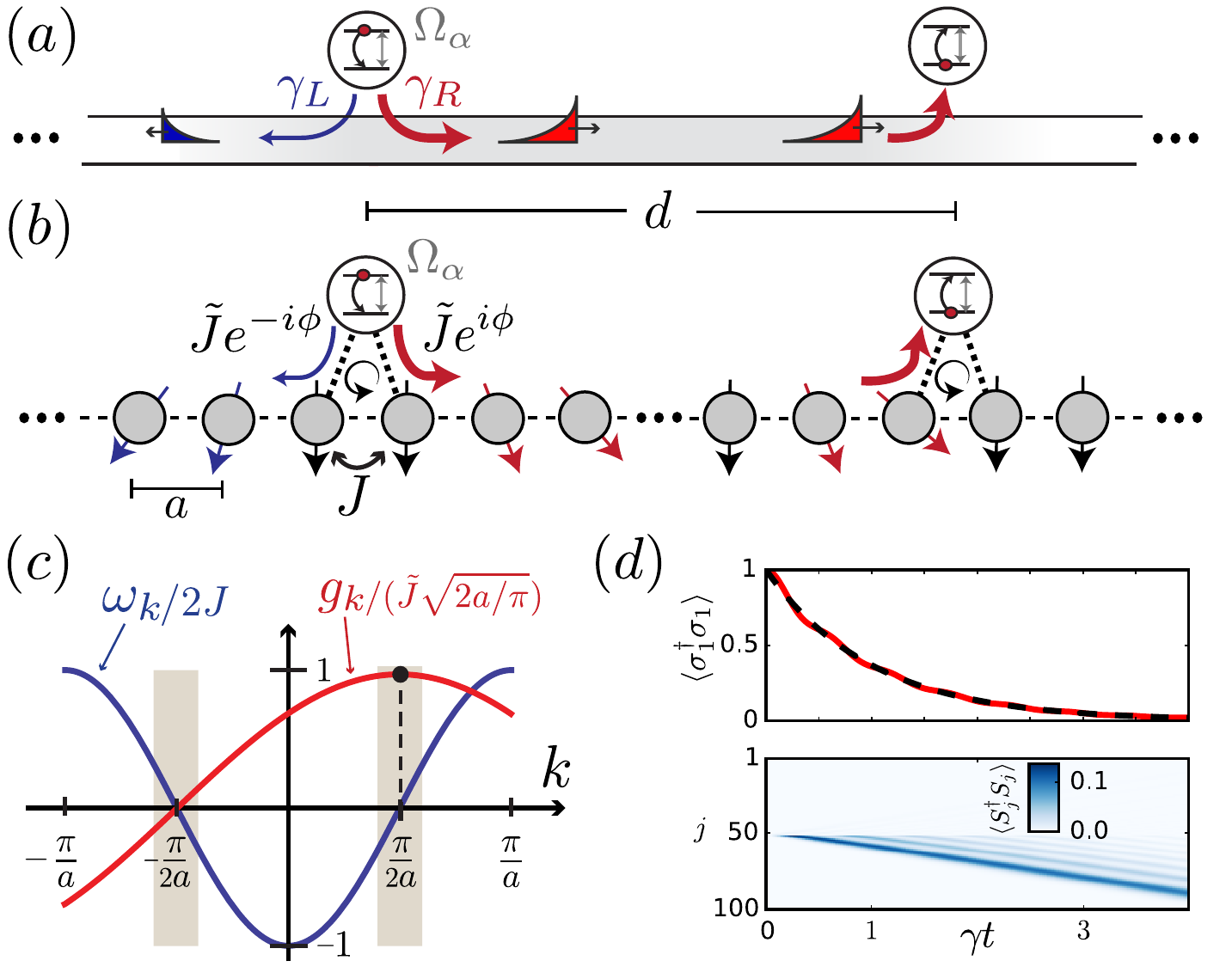}
\caption{Two-level systems or `system spins' coupled to a photonic (a) or spin waveguide (b) as basic building block of a chiral quantum network. (a) System spins decay by emitting a photon into the left- and right-moving modes of the photonic waveguide. (b) System spins decay by emitting a magnon (spin excitation) into the spin waveguide (bath) in a flip-flop process. Chirality ($\gamma_{\rm L}\neq\gamma_{\rm R}$) is achieved by introducing a complex hopping between neighboring system and bath spins. (c) Dispersion relation $\omega_k$ and asymmetric momentum coupling $g_k$. For $\tilde{\Delta}=0$ and weak coupling $(\tilde{J}/J)^2\ll 1$, the system spins couple only to resonant waveguide modes around $k=\pm\pi/2a$ (see shaded region), such that for $\phi=\pi/4$, a pure unidirectional system-bath coupling is achieved. (d) Demonstration of unidirectional emission of magnons with a single system spin coupled to the spin waveguide for $\phi=\pi/4$, $\tilde{J}=0.3J$, and $\tilde{\Delta}=0$. Dashed line corresponds to the exponential decay predicted by the Markovian theory. The lower panel shows the evolution of the bath spins occupation $\langle S_j^+ S_j^-\rangle$, evidencing the unidirectional propagation of the emitted magnons through the spin waveguide. 
\label{Sbfig:setup}}
\end{figure}

\subsection{Chiral coupling to a spin waveguide}\label{Sbsec:model}

In a photonic context a minimal building block of a chiral network is illustrated in Fig.~\ref{Sbfig:setup}(a) for $2$ two-level systems, which represent the nodes of the network. In the spirit of an {\em open} quantum system the waveguide is assumed to be {\em infinite}, providing both the communication channels between the nodes, and the input-output ports of the network [see also Fig.~\ref{Sbfig:vision}(a)]. Chirality refers to the possibility of having different coupling strengths, $\gamma_{\rm L}\ne\gamma_{\rm R}$, of the nodes to the left- and right-propagating modes in the waveguide  \cite{SbStannigel:2012jk,SbRamos2014,SbPichler2015,SbPetersen2014,SbMitsch2014,SbColes:2016dw,SbSollner2015,SbYoung2015,SbNeugebauer:2014iy,SbRodriguezFortuno:2013ew,SbLin:2013do}. Instead we consider here spin chains as communication channels [Fig.~\ref{Sbfig:setup}(b)].

Our model considers $N_\mathrm{S}$ two-level systems coupled to a spin waveguide, as illustrated in Fig.~\ref{Sbfig:setup}(b) for $N_{\rm S}=2$. Throughout this work we also refer to the two-level systems as `system spins', emphasizing that they correspond to the \emph{open system} usually considered in a Markovian network theory. These system spins have ground and excited states, $|g\rangle_\alpha$ and $|e\rangle_\alpha$, respectively ($\alpha=1,\dots,N_\mathrm{S}$), and are coherently driven with Rabi frequencies $\Omega_\alpha$ and detuning $\Delta$. As in standard quantum optics, we assume the validity of the rotating wave approximation, and we obtain in the rotating frame the following Hamiltonian for the system spins ($\hbar\equiv 1$)
\begin{align}
H_\mathrm{S}=-\Delta\sum_\alpha\sigma_\alpha^+\sigma^-_\alpha+\frac{1}{2}\sum_{\alpha} \left(\Omega_\alpha\sigma^-_\alpha+{\rm H.c.}\right)\label{SbHS}\,,
\end{align}
with $\sigma^-_\alpha\!=\!|g\rangle_\alpha\langle e|=(\sigma_\alpha^+)^\dagger$. Excitations of system spins can be transferred to the spin waveguide via dipolar flip-flop interactions. In its simplest form, the spin waveguide is realized by a large $XX$ spin chain with Hamiltonian 
\begin{align}
H_\mathrm{B}=-J\sum_{j}S_{j+1}^+ S_j^-+{\rm H.c.}\,,\label{SbHB}
\end{align}
where $S_j^-=\downket_j\upbra\ =(S_j^+)^\dagger$ is the lowering operator for a spin at site $j$ in the waveguide. For simplicity we consider here a model with nearest-neighbor (NN) hopping with amplitude $J$ \footnote{An extension of this model that includes long-range interactions in the spin waveguide can be found in the accompanying paper \cite{SbImplementation}.}. Following the same interpretation to denote the `system spins', we refer to the spins forming the waveguide as `bath spins', since they can be identified with the \emph{bath} degrees of freedom that are adiabatically eliminated in a Markovian theory. In addition, the vacuum state $|0\rangle$ of a photonic waveguide bath finds its counterpart in the spin chain prepared in the state with no spin excitations, i.e.~$|0\rangle\equiv\bigotimes_j\downket_j$.

We achieve a {\em chiral coupling} for the decay of the system spins to the spin waveguide via the coupling Hamiltonian  
\begin{align}
H_\mathrm{SB}&=\tilde{J}\sum_{\alpha}\sigma_\alpha^-\left(e^{-i\phi}S^+_{L[\alpha]}+e^{i\phi}S^+_{R[\alpha]}\right)+{\rm H.c.}\,.\label{SbHSB}
\end{align}
Here, each system spin $\alpha$ couples with strength $\tilde{J}$ to the {\em two} nearest bath spins at sites $j=L[\alpha]$ and $j=R[\alpha]=L[\alpha]+1$, located to its left and right, respectively [cf.~Fig.~\ref{Sbfig:setup}(b)]. In addition, different system spins are separated by a distance $d$ such that $L[\alpha+1]=L[\alpha]+d/a$, with $a$ the spacing between bath spins. The relative phase $\phi$ is interpreted as a synthetic gauge field \cite{SbGoldman:2014bva} that induces a flux of $2\phi$ through each of the triangular plaquettes spanned by the system--bath interactions, as shown in Fig.~\ref{Sbfig:setup}(b). The gauge field allows these interactions to imprint a {\em momentum kick} on magnons moving dominantly left or right, in close analogy to the photonic case. Complex hoppings as in Eq.\,(\ref{SbHSB}) naturally appear in the dipole-dipole interactions between  Rydberg atoms, or polar molecules~\cite{SbYao2012,SbSyzranov2014,SbPeter2015}, and we discuss details of implementing the present model system with Rydberg atoms in an accompanying paper \cite{SbImplementation}. 

\subsection{Spins vs.~photons as mediators of (chiral) interactions}\label{Sbchirality_infinite}

In the limit of small excitation probabilities $\braket{S_j^+S_j^-}\ll 1$, we can use spin-wave theory~\cite{SbDiep2004} to bosonize the bath spin excitations $S_j^-\rightarrow b_j$, with $[b_j,b_l^\dag]=\delta_{jl}$. For an infinitely long chain, the Hamiltonian in Eq.\ (\ref{SbHB}) then resembles a photonic bandgap material \cite{SbJohn1990,SbDouglas2015},
\begin{align}
H_\mathrm{B}&=\int dk\ \omega_k b^\dag_k b_k.\label{SbHBbulk}
\end{align}
Here, the delocalized bosonic operators $b_k=(1/2\pi)^{1/2}\sum_j b_j e^{-ikaj}$ annihilate a magnon excitation in the waveguide with quasi-momentum $k\in[-\pi/a,\pi/a]$ and nonlinear dispersion $\omega_k=-2J\cos(ka)$. As illustrated in Fig.~\ref{Sbfig:setup}(c) for $J>0$, magnons with $k>0$ ($k<0$) have positive (negative) group velocity $v_k=\partial\omega_k/\partial k$, and thus propagate to the right (left) along the spin waveguide \footnote{Since interchanging $J\leftrightarrow-J$ simply reverses the definition of left and right movers, we focus in the following on the case $J>0$ without loss of generality.}. 

Writing the flip-flop coupling Hamiltonian (\ref{SbHSB}) also in terms of the delocalized bosonic modes $b_k$, we obtain a standard quantum-optical system-bath interaction \cite{Sbgardiner2014,Sbgardiner2015,SbLehmberg:1970jj,SbGonzalezTudela:2013hn},
\begin{align}
H_\mathrm{SB}&=\sum_{\alpha}\int dk\ g_k e^{-i\alpha kd}\sigma_\alpha^- b^\dag_k+{\rm H.c.},\label{SbHSBbulk}
\end{align}
but with an engineered momentum-dependent coupling given by
\begin{align}
g_k&=\tilde{J}\sqrt{2a/\pi}\cos(ka/2-\phi).\label{Sbgk}
\end{align}
Importantly, the phase $\phi$ renders this coupling asymmetric in $k$ and thus makes it chiral. Figure~\ref{Sbfig:setup}(c) illustrates chirality for $\phi\!=\!\pi/4$, where all waveguide excitations moving to the right couple stronger than the ones to the left. When considering the modes with $k\!=\!\pm\pi/2a$, in particular, perfect unidirectionality is achieved. Since changing $\phi\!\leftrightarrow\! -\phi$ in Eq.~(\ref{Sbgk}) merely reverses the preferred coupling between $k\!\leftrightarrow \!-k$, we assume without loss of generality the convention $\phi\in[0,\pi/4]$, allowing us to tune the chirality from bidirectional ($\phi=0$) to perfectly unidirectional to the right ($\phi=\pi/4$). 

While in this section we have made a formal identification of the spin waveguide with a structured reservoir of non-interacting photons (provided $\langle S_j^+S_j^-\rangle\ll 1$), we emphasize that our model in general includes the interactions due to the hard-core nature of the magnons, which we investigate in Secs.~\ref{Sbsec:NonMark}-\ref{Sbsec:spintronics}.

\subsection{Markovian theory of chiral quantum networks: master equation for system spins} \label{Sbsec:markov}

We can eliminate the spin waveguide as a quantum bath in the Born-Markov approximation to derive a ME for the dynamics of the systems spins. This is valid under the following assumptions: (i) weak system-bath coupling $(\tilde{J}/J)^2\ll 1$, (ii) negligible coupling to modes at the band-edge $|\Delta|/2J\ll 1$, and (iii) negligible propagation time of the magnons compared to the relevant timescales of the system spin dynamics \cite{SbLehmberg:1970jj,SbMiloni1974,SbChang:2012co}. 

As detailed in Appendix \ref{Sbapp:infiniteMasterChiral}, the result of this adiabatic elimination of the spin waveguide is the chiral master equation \cite{SbRamos2014,SbPichler2015}   
\begin{align}
\dot{\rho}_\mathrm{S}=-i[H_\mathrm{S}+H_\mathrm{C},\rho_\mathrm{S}]+\gamma_\mathrm{R}{\cal D}[c_\mathrm{R}]\rho_\mathrm{S}+\gamma_{\mathrm{L}}{\cal D}[c_{\mathrm{L}}]\rho_\mathrm{S},\label{SbchiralMaster}
\end{align}
with $\rho_\mathrm{S}$ the reduced density operator of the system spins (nodes). As a consequence of the chiral coupling, the decay rates $\gamma_L$ and $\gamma_R$ into left (L)- and right (R)-moving magnons, respectively, are in general asymmetric and read
\begin{align}
\gamma_\nu&=2\pi \frac{g_{\nu\bar{k}}^2}{|v_{\bar{k}}|}=\frac{\tilde{J}^2}{J}\frac{[1\!+\!\cos(\bar{k}a-2\nu\phi)]}{\sin(\bar{k}a)},\label{SbdecayGeneral}
\end{align}
where we assigned the values $\nu=\{+1,-1\}$, corresponding to $\nu=\{R,L\}$, and $\bar{k}a=\arccos(\tilde{\Delta}/2J)>0$ is the resonant wavevector of the right moving waveguide excitations. Notice that the detuning is renormalized by a Lamb shift as $\tilde{\Delta}=\Delta-(\tilde{J}^2/J)\cos(2\phi)$ [cf.~Appendix \ref{Sbapp:infiniteMasterChiral}]. Throughout this work we are mainly interested in the case $\tilde{\Delta}=0$, where the system spins decay resonantly into waveguide excitations with $k=\pm\pi/2a$ and group velocity $\bar{v}=\pm 2Ja$ [cf.~Fig.~\ref{Sbfig:setup}(c)]. The corresponding decay rates reduce simply to $\gamma_\nu=(\gamma/2)[1+\nu\sin(2\phi)]$, such that for $\phi=\pi/4$ we obtain a perfect unidirectional coupling with $\gamma_{\rm L}=0$ and $\gamma_{\rm R}=\gamma$, where $\gamma=\gamma_{\rm R}+\gamma_{\rm L}=2\tilde{J}^2/J$ is the total decay rate. In addition, the chiral Hamiltonian $H_\mathrm{C}$ describes infinite-range reservoir-mediated interactions between the system spins
\begin{align}
H_\mathrm{C}=\frac{i}{2}\sum_{\alpha<\beta}\!\left(\gamma_{\rm R} e^{-i\varphi_{\alpha\beta}}-\gamma_{\rm L} e^{i\varphi_{\alpha\beta}}\right)\sigma^+_\alpha\sigma^-_\beta+{\rm H.c.},\label{SbchiralH}
\end{align}
with $\varphi_{\alpha\beta}=\bar{k}d|\alpha-\beta|$, phases accumulated due to the magnon propagation. The Lindblad operator ${\cal D}[A]\rho=A\rho A^\dag-(A^\dag A\rho+\rho A^\dag A)/2$ describes the Markovian dissipative processes with collective left and right jump operators given by
\begin{align}
c_{\nu}=\sum_\alpha e^{-i\nu\alpha \bar{k}d}\sigma^-_\alpha.\label{SbglobalJump}
\end{align}
These collective jump operators are also obtained in the Dicke model of superradiance for isotropic baths \cite{SbDicke:1954bl,SbLehmberg:1970jj}, but here we have the additional feature of tuning the directionality of the emission into the waveguide via the phase $\phi$. For instance, when $\phi=\pi/4$ the system spins behave as cascaded quantum systems \cite{SbGardiner1993,SbCarmichael1993,Sbgardiner2015}, whereas when $\phi=0$ we recover the one-dimensional Dicke model for a bidirectional bath \cite{SbChang:2012co,SbGonzalezTudela:2013hn,SbGoban:2015dr}. We remark that the master equation derived here is formally identical to the one of the {\em Markovian chiral photonic network model} \cite{SbRamos2014,SbPichler2015}, and the corresponding results for quantum many-body dynamics and quantum information protocols carry over to the chiral quantum network with spin waveguides in the weak-coupling Markovian limit. For instance, the Markovian exponential decay of a single system spin and its unidirectional emission into the spin waveguide is numerically demonstrated in Fig.~\ref{Sbfig:setup}(d).  

\subsection{Beyond the Born-Markov approximation: the extended master equation}\label{Sbsub:beyondMarkExt}

In order to go beyond the Born-Markov approximation, the dynamics of the spin waveguide has to be included on the same footing as the system spins (nodes). Since an exact treatment of the infinite waveguides is computationally intractable, we make use of the fact that spin excitations leaving the network do not return. Thus we define an \emph{extended Markovian cut} keeping the nodes and the connecting spin waveguides as our Markovian 'extended system', while treating the part of the waveguides representing the input/output channels of the network as Markovian reservoirs [cf.~Fig.~\ref{Sbfig:vision}(b)]. 

\begin{figure}[t]
\includegraphics[width=0.5\textwidth]{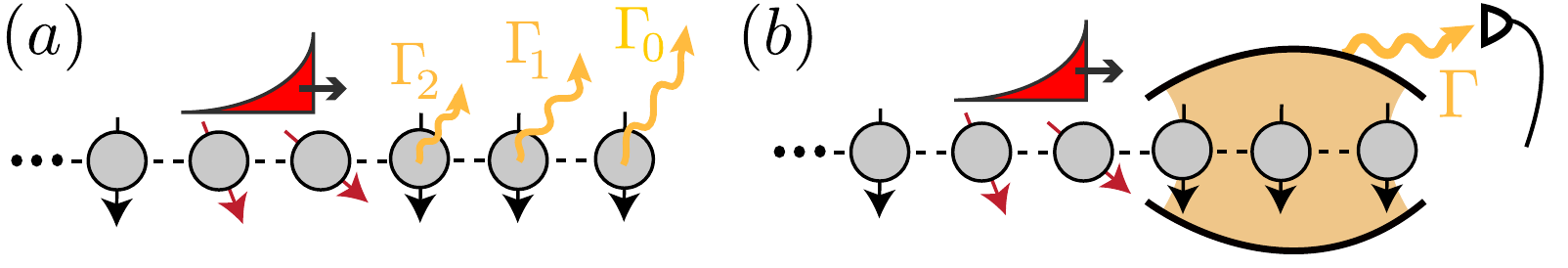}
\caption{A spin -- optical interface converting right propagating spin excitations into optical photons propagating in a fiber. (a) The inclusion of such a finite waveguide with local losses at its ends, in the calculation of the network dynamics, extends the standard Markovian cut of Quantum Optics. (b) Possible experimental realization of the spin chain losses by a spin-photon interface obtained by coupling the edge bath spins to a cavity, decaying in an optical fiber.}
\label{Sbfig:markovian_cut}
\end{figure}

On a technical level, this is realized by representing the infinite waveguide by a finite chain of $N_{\rm B}$ bath spins, with absorbing boundary conditions~\cite{SbGivoli:1991js}. To this end, we introduce local losses $\Gamma_n$ on $M$ bath spins at each end of the chain, and increase the loss rates smoothly towards the boundary in order to minimize reflections [cf.~Fig.~\ref{Sbfig:markovian_cut}(a)]. The full dynamics of the relevant part of the network is then described by an {\em extended Markovian master equation} as
\begin{align}
\dot{\rho}=-i[H,\rho]+\!\sum_{n=0}^{M-1}\!\Gamma_{n}\left({\cal D}[S^-_{1+n}]\rho+{\cal D}[S^-_{N_{\rm B}-n}]\rho\right),\label{SbfullMaster}
\end{align}
where $\rho(t)$ is the density matrix of the system spins plus the finite spin waveguide with sites $j=1,\dots,N_{\rm B}$, $H=H_\mathrm{S}+H_\mathrm{B}+H_\mathrm{SB}$ is the corresponding total Hamiltonian, and $\Gamma_n$ is the decay rate of bath spins at $n$ sites away from the ends of the chain. Throughout this work we consider only one loss per end with an optimized decay of $\Gamma_0=2J$, which is sufficient to engineer absorbing boundaries with negligible reflections. This condition becomes exact in the weak coupling limit, as shown in Appendix~\ref{Sbapp:FiniteMaster} \footnote{In the case of a waveguide with long-range and/or inhomogeneous couplings, many smoothly increasing losses are in general needed, as discussed in Ref.~\cite{SbImplementation}.}.

Importantly, by solving the extended ME (\ref{SbfullMaster}) we model the non-Markovian dynamics of the quantum network. This is in the spirit of representing non-Markovian quantum stochastic processes as {\em projection} of a quantum Markov process for an extended number of degrees of freedom, known in the literature as a Markovian embedding \cite{SbBreuer:2004fc,SbWoods:2014kn,SbHughes2009}. In addition, this extended ME can be solved efficiently using matrix product states (MPS) techniques \cite{SbSchollwock2011}, adapted to describe the evolution of open quantum systems \cite{SbVerstraete:2004hi,SbDaley2014,SbWerner:2014wq,SbCui:2015im,SbMascarenhas:2015cc}. In this work we use a quantum trajectories approach \cite{SbDaley2014}, and obtain the dynamics in Eq.~(\ref{SbfullMaster}) by a stochastic average over many independent evolutions or quantum trajectories. Each trajectory $m$ is represented by a MPS $|\Psi_m(t)\rangle$, whose dynamics is governed by a non-Hermitian Hamiltonian and quantum jumps occurring with a probability given by the loss rates $\Gamma_n$ \cite{SbDaley2014}. MPS techniques have been originally developed in a condensed matter context to enable an efficient description of many-body quantum states in 1D systems and to integrate the many-particle Schr\"odinger equation. In the present open spin network context, it allows us to access the regime of long spin chains with multiple excitations propagating {\em in} the waveguide (including system-bath entanglement), constituting a situation where the exact representation becomes inefficient.

In Ref.~\cite{SbPichler2015b}, an alternative approach to problems with long time delays was introduced. The method developed there is tailored to describe situations where the non-Markovian dynamics stems solely from time delays of photons propagating between the network nodes while the coupling of each individual node to the waveguide is treated in a Markovian approximation. It is based on a linear photon dispersion relation allowing for a transparent formulation of the problem in a time bin basis such that the photon propagation is already accounted for by the formalism, simplifying the MPS description. For the spin waveguides considered here, all the interactions are local in a real space representation, allowing us to also account for dispersion effects due to a strong system-bath coupling, in addition to the possibly long time delays in the propagation of magnons and the collisions between them. Other works were MPS techniques have been used to efficiently simulate the dynamics of discretized 1D waveguides can be found in Refs.~\cite{SbPeropadre:2013ia,SbPrior:2010gd}, where the spin-boson model in the strong coupling regime is considered. An extension of the MPS treatment to efficiently describe thermal baths is developed in Ref.~\cite{SbdeVega:2015dm}.

\subsection{Spin waveguide - optical waveguide interface}

In an open quantum network with infinite spin waveguides the {\em extended Markovian cut} [cf.~Fig.~\ref{Sbfig:vision}(b)] introduced above has a direct physical meaning. A physical spin waveguide will never be infinite, but always be terminated. It therefore seems natural and useful to assume that the spin-wave excitations will be converted (without reflection) via a spin-optical interface into, for example, optical photons propagating in a fiber, as outlined in Fig.~\ref{Sbfig:markovian_cut}(b). 

\section{Non-Markovian dynamics of chiral quantum networks}\label{Sbsec:NonMark}

In this section we study driven-dissipative dynamics of chiral quantum networks of system spins coupled to the spin waveguide beyond the Markovian regime. As discussed in Sec.~\ref{Sbsub:transient} below, the sources of non-Markovianity are the non-linear Bloch band dispersion relation, time delays in the magnon propagation, and the possibility of a strong system-bath coupling. The framework to describe the non-Markovian dynamics is provided by the extended ME introduced in Sec.~\ref{Sbsec:CQN}, which keeps the full dynamics of the bath excitations in the spin waveguides connecting the driven system spins, while moving the Markovian cut to the input/output ports of the network.  

Sec.~\ref{Sbsub:transient} below discusses non-Markovian effects in transient dynamics for decay of driven system spins to the spin chain, and their characterization using recently introduced witnesses of non-Markovianity~\cite{SbRivas2014,SbBreuer2015}. Sec.~\ref{Sbsub:nonMarksteady} illustrates non-Markovianity in the steady state regime. We choose as example the formation of quantum dimers, recently discussed for driven-dissipative Markovian chiral quantum networks \cite{SbStannigel:2012jk,SbRamos2014,SbPichler2015}, and we characterize the non-Markovianity in steady-state via two-time correlations, quantum mutual information and entanglement entropy, demonstrating the fundamentally different behavior of unidirectional ($\gamma_{\rm L}=0$) vs.~asymmetric bidirectional systems ($\gamma_{\rm R}>\gamma_{\rm L}\neq 0$) subjected to time delays.

Although the chiral network model is given in Eqs.~(\ref{SbHS})-(\ref{SbHSB}) for a spin waveguide, we stress that the analysis below also includes the case of a structured photonic waveguide, obtained by replacing $S_j^-\to b_j$. In the low occupation limit $\langle S_j^+S_j^-\rangle\ll 1$ both cases are equivalent, as shown in Sec.~\ref{Sbchirality_infinite}, so we only consider the spin case explicitly. In contrast, for $\langle S_j^+S_j^-\rangle\gtrsim 1$, as in Sec.~\ref{Sbsub:hardcore}, the behavior of a spin and a photon waveguide qualitatively deviate from each other, so in that section we explicitly calculate the dynamics for both cases and compare them.

\subsection{Transient dynamics in the non-Markovian regime}\label{Sbsub:transient}

In the following we identify the mechanism behind non-Markovian effects in our networks with spin waveguides, and analyze their consequences in the transient dynamics. In Sec.~\ref{Sbsub:bandedge}, we consider effects due to the finite width of the Bloch band in competition with strong coupling of the system spins to the spin waveguide ($\gamma\sim 2J$), while in Sec.~\ref{Sbsub:retardation} we discuss retardation effects due to a finite propagation time $\tau=d/|\bar{v}|$ of magnons between nodes ($\gamma\tau\gg 1$). We are particularly interested in the case of strong driving ($|\Omega_\alpha|\sim\gamma$), which goes fundamentally beyond the Born-Markov approximation, and show how the network dynamics can be efficiently simulated with MPS methods and quantum trajectories. In Sec.~\ref{Sbsub:hardcore} we consider the extreme case of very long delay lines between nodes, where many excitations populating the spin waveguide collide, and thus evidence their hard-code nature in contrast to the more familiar case of (non-interacting) photons as mediators of interactions.

\subsubsection{Band-edge effects in the presence of strong driving}\label{Sbsub:bandedge}

The lattice structure inherent to the spin waveguide results in a Bloch-band dispersion for magnons as depicted in Fig.~\ref{Sbfig:setup}(c). In the strong coupling limit ($\gamma\sim 2J$), system spins unavoidably couple to the whole band, and in particular to the modes at the edges ($k=0,\pm \pi/a$), which do not propagate due to a vanishing group velocity.

In the absence of driving ($\Omega=0$), the corresponding dynamics can be studied analytically by a Wigner-Weisskopf (WW) ansatz, describing the dynamics of a single spin excitation in the quantum network [cf.~Appendix \ref{Sbapp:WW} for details]. The main feature of the coupling to the band-edge is the presence of one or two {\em  bound states} (for details see Refs.~\cite{SbJohn1990,SbKofman1994,SbNavarrete-Benlloch2011,SbDouglas2015,SbCalajo2015,SbLombardo:2014ju,SbSanchezBurillo:2016vg}), which are a superposition of a system spin excitation and a localized contribution of waveguide excitations,
$|\Psi\rangle_{\rm bound}=(c_1\sigma^+_1+\sum_j \bar{c}_j S_j^+) |g\rangle|0\rangle$, with an eigenenergy $\omega_\mathrm{BS}$ outside the waveguide dispersion relation ($|\omega_\mathrm{BS}|>2J$).

The existence of the bound states invalidates the Markovian exponential decay: a system spin does not decay completely into the waveguide, as a fraction of the excitation, corresponding to the bound-state contribution, remains trapped for infinitely long times around the position of the emitter. As an illustration, let us consider an initially excited system spin, with resonant ($\Delta=0$) and unidirectional ($\phi=\pi/4$) coupling. For $Jt\gg 1$, we can neglect the non-analytic part of the dynamics [cf.~Appendix~\ref{Sbsubsub:bandedge}], such that the excitation probability of the initially ($t=0$) excited system spin evolves as
\begin{equation}
\langle\sigma^+_1\sigma^-_1\rangle(t)=\left|\frac{(\lambda+1)}{2\lambda}e^{-\bar{\gamma}t}+\frac{(\lambda-1)}{\lambda}\cos\left(\omega_\mathrm{BS}t\right)\right|^2,\label{Sbeq:boundstate}
\end{equation}
with $\lambda=\sqrt{1+(\tilde{J}/J)^2}$ determined by an arbitrarily strong system-bath coupling. In addition to exponential decay with modified rate $\bar{\gamma}=\sqrt{2(\lambda-1)}J$, familiar from the Markov approximation, the population of the system spin oscillates with frequency $\omega_\mathrm{BS}=\sqrt{2(\lambda+1)}J>2J$ due to the coupling to two bound-states with energies $\pm\omega_\mathrm{BS}$. In the regime of small couplings, $(\tilde{J}/J)^2\ll 1$, the bound-state contribution can be neglected whereas in the limit of large couplings, the oscillation of the two-bound states is dominant.

\begin{figure}
\includegraphics[width=\columnwidth]{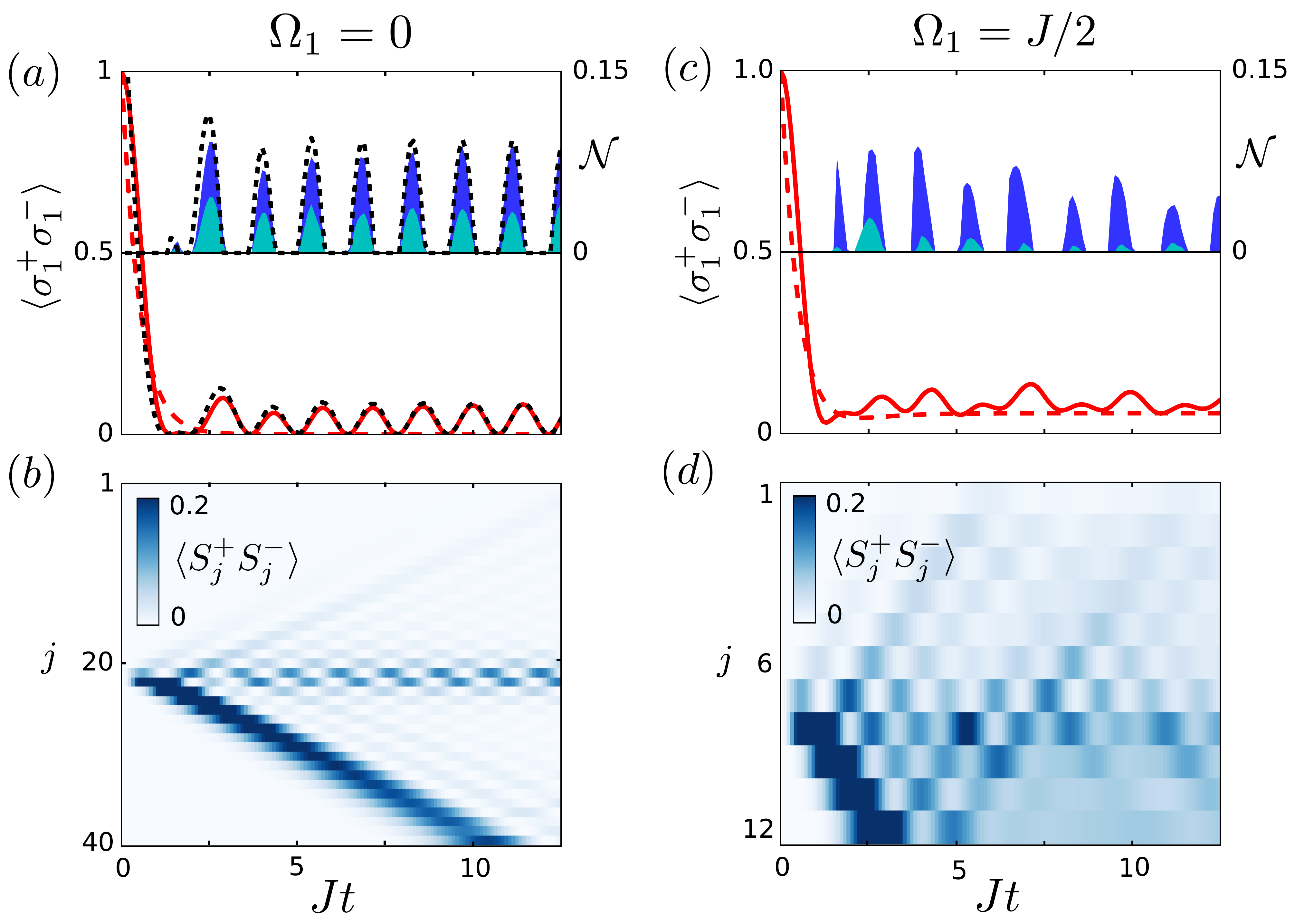}
\caption{Band-edge effects in the dynamics of a single strongly coupled system spin, in a absence (a,b) and presence (c,d) of driving. (a) Occupation $\langle \sigma_1^+\sigma_1^-\rangle$ of an undriven ($\Omega_1=0$) and strongly coupled ($\tilde{J}/J=1$) system spin, showing a fast decay and a permanent oscillation due to the coupling to two bound states when $\phi=\pi/4$ and $\Delta=0$ [cf.~Appendix \ref{Sbsubsub:bandedge}]. The Markovian and WW predictions are shown as a dashed and dotted line, respectively. The non-Markovian behavior is detected via non-zero values of the BLP and RHP witnesses, represented by the blue and cyan regions, respectively. The BLP witness is calculated by comparing the evolution from the initial conditions $\rho^{(1)}_{\rm S}=|e\rangle\langle e|$ and $\rho^{(2)}_{\rm S}=|g\rangle\langle g|$, while the RHP witness is obtained by initially preparing the system and ancilla in the entangled state $\rho_{\rm SA}(0)=|\Psi_0\rangle_{\rm SA}\langle \Psi_0|$, with $|\Psi_0\rangle_{\rm SA}=(|e\rangle_{\rm S}|e\rangle_{\rm  A}+|g\rangle_{\rm  S}|g\rangle_{\rm A})/\sqrt{2}$. The WW prediction for the BLP witness is represented by a dotted line. (b) For the same situation as in (a), we plot the evolution of the bath spins occupation $\langle S_j^+ S_j^-\rangle$, showing the localized bound state oscillation around the system spin position. In addition, we observe a nearly unidirectional propagation of the magnons during the decay of the system spin. Despite that $\phi=\pi/4$, there is also a small fraction of magnons propagating to the left as the strong coupling allows the system spin to decay to modes different than $k=\pm\pi/2a$ [cf.~Fig.~\ref{Sbfig:setup}(c)]. (c,d) Dynamics of system and bath spins, under the same conditions as in (a,b), but with the inclusion of driving $\Omega_1=\gamma/4$ in the system. The driving tends to wash out the band-edge oscillations, but they are still identified by the witnesses as the dominant non-Markovian behavior.
\label{Sbfig:NonMark_time_evo}}
\end{figure}

The techniques introduced in Sec.~\ref{Sbsub:beyondMarkExt} allow us to go beyond the WW description, and include a strong coherent drive $|\Omega_1|\sim\gamma\sim J$, which populates the waveguide with many excitations in addition to the bound states. In Fig.~\ref{Sbfig:NonMark_time_evo}, we show the dynamics of the system and waveguide occupations in the strongly coupled regime, for the driven and undriven case. After a time $t\gtrsim 1/\bar{\gamma}$, the system spin population differs significantly from the Markovian approximation due the bound-state contribution that permanently exchanges occupation between system and bath spins [cf.~Fig.~\ref{Sbfig:NonMark_time_evo}(a,c)]. As shown in Fig.~\ref{Sbfig:NonMark_time_evo}(b,d), the chiral emission into the waveguide, as well as the localized bound-state oscillation can be directly observed in the waveguide dynamics. In the case of strong driving, the continuous emission of magnons tends to damp out the system spin oscillations.

To quantify the degree of non-Markovianity for general open system dynamics, a set of witnesses and measures have been proposed in recent literature~\cite{SbRivas2014,SbBreuer2015}. We complement our discussion by considering two of these non-Markovian witnesses, namely the Breuer-Laine-Piilo (BLP) and the Rivas-Huelga-Plenio (RHP), which provide us with criteria to exclude a potential Markovian theory that can describe the reduced evolution of the open system state $\rho_{\rm S}(t)={\rm Tr}_{\rm B}\lbrace\rho(t)\rbrace$, when tracing over the bath spins (waveguide). 

The BLP witness~\cite{SbBreuer2009} compares the evolution for various initial conditions, $\rho^{(1)}_{\rm S}(0)$ and $\rho^{(2)}_{\rm S}(0)$, and  quantifies the distinguishability of the resulting density matrices through the trace distance as $\mathcal{D}(t)=||\rho_\mathrm{S}^{(1)}(t)-\rho_\mathrm{S}^{(2)}(t)||/2$, with $||A||={\rm Tr}(\sqrt{A^\dag A})$. An increase of this distinguishability in time is interpreted as an information back-flow from the spin waveguide (bath) to the system spins (open system $S$), as a hallmark of non-Markovian behavior. Whereas the construction of a non-Markovian measure requires a maximization over all initial conditions, a practical lower bound is obtained by calculating ${\cal D}(t)$ for only two well-chosen initial states. Non-Markovianity is then witnessed when the quantity \begin{align}
{\cal N}_{\rm BLP}(t)=\frac{d{\cal D}}{dt}\ \Theta\left(\frac{d{\cal D}}{dt}\right),
\end{align}
is nonzero, with $\Theta(t)$ the Heaviside function. 

Plenio and collaborators introduced a complementary non-Markovian witness, where Markovian dynamics is identified with completely-positive trace-preserving evolutions (CPT)~\cite{SbRivas2010}. The recipe proposed for extracting non-Markovianity is to consider an ancillary copy $\rm A$ of the open system $\rm S$, with A and $\rm S$ prepared initially in a maximally entangled state $\rho_{\rm SA}(0)$. Subsequently, the open system $\rm S$ evolves due to the coupling to the bath, while the ancilla is kept isolated. Under CPT evolutions, the entanglement between system and ancilla can only decrease. Thus, an increase in time of any entanglement monotone $E[\rho_{\rm SA}(t)]$ (as for instance the negativity used here~\cite{SbVidal2002}) witnesses non-Markovianity. Although with this criterion it is also possible to construct a non-Markovian measure by using in addition process tomography, for the present discussion it is sufficient to consider only the experimentally more accessible witness defined as
\begin{align}
{\cal N}_{\rm RHP}(t)=\frac{dE[\rho_{\rm SA}(t)]}{dt}\ \Theta\left(\frac{dE[\rho_{\rm SA}(t)]}{dt}\right),
\end{align} 
where $\rho_{\rm SA}(0)$ is a well-chosen entangled state, but not necessarily maximally entangled.
  
In Fig.~\ref{Sbfig:NonMark_time_evo}(a,c), we display ${\cal N}_{\rm BLP}(t)$ and ${\cal N}_{\rm RHP}(t)$ for the above case of a system spin that is strongly coupled to an unidirectional waveguide. The two witnesses agree well in identifying the bound-state oscillations as the dominant non-Markovian behavior. This is expected in this single emitter case, since both witnesses depend on the time derivative of the system spin occupation. Related works where the dynamics of bound-states in bidirectional waveguides have been studied can be found in \cite{SbLombardo:2014ju,SbSanchezBurillo:2016vg}.

\subsubsection{Retardation effects in the presence of strong driving}\label{Sbsub:retardation}
   
A quantum network consists in general of many nodes, which communicate over a distance via emission and absorption of waveguide excitations. The finite propagation speed of these excitations naturally introduces time delays in the dynamics, which can invalidate the Markov approximation.

The minimal model to study non-Markovian effects of retardation is given by two resonantly driven system spins ($\tilde{\Delta}=0$), with chiral coupling to a spin waveguide, and separated by a large distance $d$ such that the delay between them is large compared to their lifetime, $\tau=d/|\bar{v}|\gg 1/\gamma$. Again, we can understand some basic aspects of this problem by first considering the undriven case ($\Omega_\alpha=0$) within a WW approach [cf.~Appendix \ref{Sbsubsub:retardation}]. In the weak coupling limit ($\gamma\ll 2J$), and assuming that only the left system spin $\alpha=1$ is initially excited, the occupations of the system spins evolve as
\begin{align}
\langle \sigma^+_1\sigma^-_1\rangle(t)&=\left|\sum_{n=0}^\infty\gamma_{\rm L}^{n}\gamma_{\rm R}^n f^{(0)}_n(t-2n\tau)\right|^2,\label{SbWWret1}\\
\langle \sigma^+_2\sigma^-_2\rangle(t)&=\left|\sum_{n=0}^\infty \gamma_{\rm L}^{n}\gamma_{\rm R}^{n+1} f^{(1)}_n(t-[2n+1]\tau)\right|^2,\label{SbWWret2}
\end{align}
with $f_{n}^{(m)}(t)=e^{i\bar{k}d(2n+m)}t^{(2n+m)}e^{-\gamma t/2}\Theta(t)/(2n+m)!$. As described by the functions $f_n^{(m)}$ in Eqs.~\eqref{SbWWret1}-\eqref{SbWWret2}, the dynamics consists of a succession of emission and re-absorption of the waveguide excitation from spin $\alpha=1$ to $\alpha=2$ and vice versa, occurring at times that are multiples of the magnon propagation time $\tau$. Equations~\eqref{SbWWret1}-\eqref{SbWWret2} generalize the results of Milonni and Knight \cite{SbMiloni1974,SbDung:1999cj} to a chiral 1D channel with possibly asymmetric decays $\gamma_{\rm L}\leq\gamma_{\rm R}$. This directional coupling can strongly influence the cycles of emission and re-absorption, as it is apparent in the extreme unidirectional limit ($\gamma_L=0$), where only one excitation transfer from left to right is possible. 

When adding a strong drive to the nodes ($|\Omega_\alpha|\sim\gamma$), the non-Markovian regime of large retardation $\gamma\tau\gg 1$ can no longer be treated analytically due to the many emitted magnons that propagate through the waveguide before they are reabsorbed by the other node. Nevertheless, our extended ME method (\ref{SbfullMaster}) allows us to describe the system-bath dynamics also in this non-trivial regime, and in particular, to reach the steady state.

\begin{figure}[t!]
\includegraphics[width=\columnwidth]{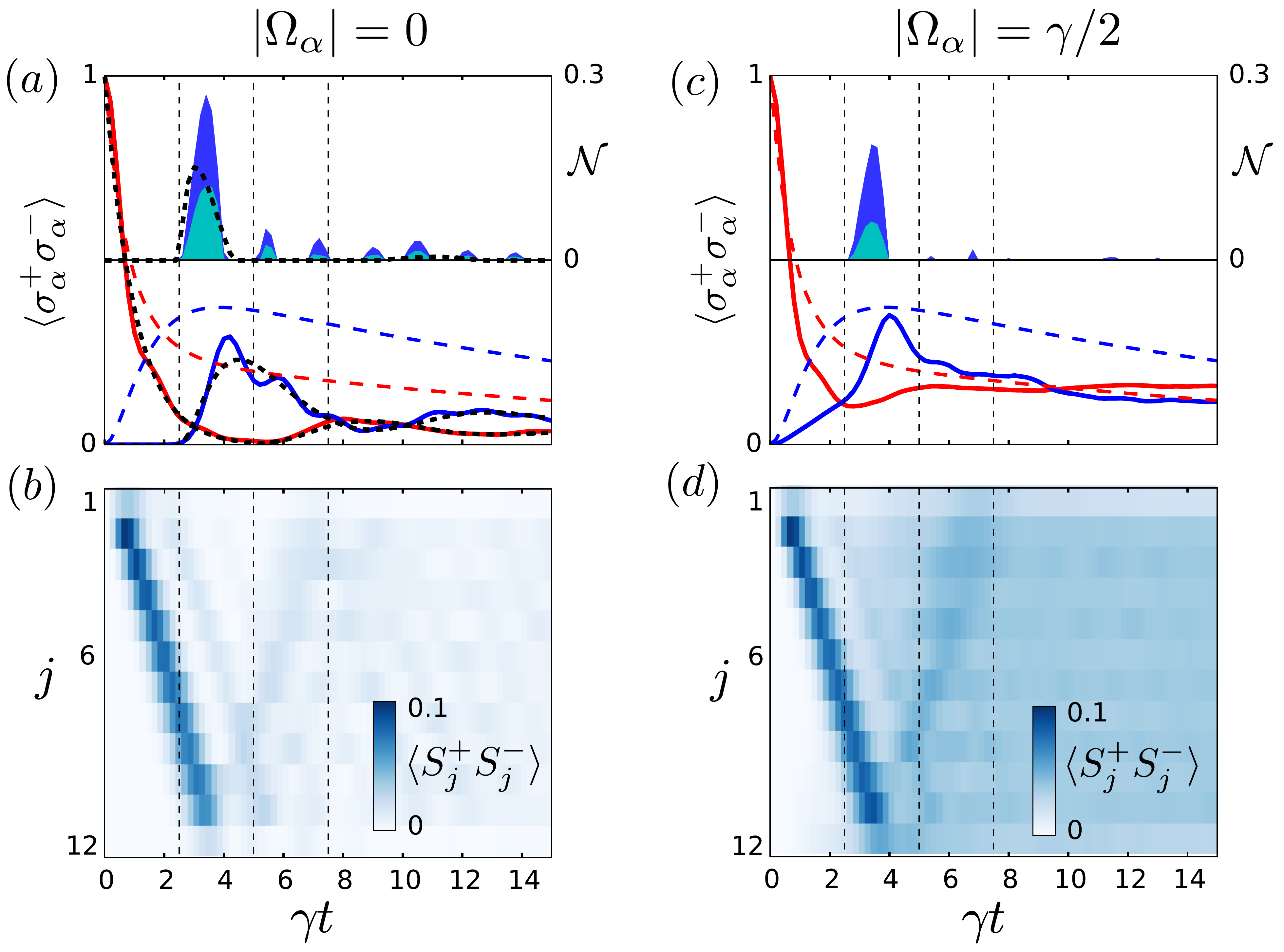}
\caption{Retardation effects in the dynamics of two distant system spins, in the absence (a,b) and presence (c,d) of driving. (a) Occupations $\langle \sigma_1^+\sigma_1^-\rangle$ (red) and $\langle \sigma_2^+\sigma_2^-\rangle$ (blue) of the system spins sitting on the left and right end of the waveguide, respectively. They are separated by a distance $d=10a$, and are chirally coupled to the spin waveguide with $\tilde{J}=0.5J$, $\phi=\pi/10$, and $\tilde{\Delta}=0$ (such that $\tau=2.5/\gamma$ and $\gamma_{\rm L}/\gamma_{\rm R}\approx0.26$). We observe the chiral decay of the initially excited left system spin, and the retarded excitation of the right system spin at $t\approx\tau$. This is well-described by the WW prediction (dotted lines), but it strongly deviates from the Markovian prediction (dashed lines) as the reabsorptions are assumed to be instantaneous. Small deviations from the WW prediction stem from a weak coupling to the band-edge modes, which are not accounted by the analytical expressions (\ref{SbWWret1})-(\ref{SbWWret2}) [cf.~Appendix \ref{Sbsubsub:retardation}]. The non-Markovian witnesses BLP (blue) and RHP (cyan) identify the absorptions at times multiples of $\tau$, as well as the residual oscillations due to band-edge effects. The two initial conditions used to calculate the BLP witness are $\rho^{(1)}_{\rm S}=|eg\rangle\langle eg|$ and $\rho^{(2)}_{\rm S}=|gg\rangle\langle gg|$, whereas the RHP witness assumes an initial entangled state between system and ancilla given by $\rho_{\rm SA}(0)=|\Psi_0\rangle_{\rm SA}\langle \Psi_0|$, with $|\Psi_0\rangle_{\rm SA}=(|eg\rangle_{\rm S}|eg\rangle_{\rm A}+|gg\rangle_{\rm S}|gg\rangle_{\rm A})/\sqrt{2}$. (b) For the same situation as in (a), we plot the evolution of the bath spins occupation $\langle S_j^+ S_j^-\rangle$, showing that the first emitted magnon reaches the right system spin at $t\approx\tau$, when it is absorbed and chirally reemitted. (c,d) Dynamics of system and bath spins, under the same conditions as in (a,b), but including driving on both system spins ($\Omega_1=-\Omega_2=\gamma/2$). In this situation the analytical WW treatment cannot be applied, but the retarded absorptions are clearly identified as the dominant non-Markovian behavior.}
\label{Sbfig:NonMark_time_evo_O}
\end{figure}

As an example, we show in Fig.~\ref{Sbfig:NonMark_time_evo_O} the evolution of the occupations of the two system spins as well as the spin waveguide for a considerable delay of $\gamma\tau=2.5$ and chirality $\gamma_{\rm L}/\gamma_{\rm R}\approx 0.26$. The complete cycle of decay of the initially excited system spin $\alpha=1$, the directional propagation of the emitted magnon through the waveguide, and the subsequent absorption by the second system spin $\alpha=2$ at the retarded time $t=\tau$, is clearly visible for the undriven [cf.~Fig.~\ref{Sbfig:NonMark_time_evo_O}(a,b)] and driven [cf.~Fig.~\ref{Sbfig:NonMark_time_evo_O}(c,d)] cases. The Markovian prediction from Eq.~\eqref{SbchiralMaster} highly deviates from this behavior, as it assumes this process to occur instantaneously (dashed lines). This is also consistent with the non-Markovian witnesses ${\cal N}_{\rm BLP}$ and ${\cal N}_{\rm RHP}$, introduced in the previous subsection, which identify this retarded absorption at $t=\tau$ as the largest non-Markovian aspect in the dynamics. Other smaller peaks in the non-Markovian witnesses correspond to higher-order emission and reabsorption cycles as well as residual band-edge oscillations caused by the system-bath coupling of intermediate strength $\tilde{J}=0.5J$. These two sources of non-Markovianity can be discriminated by comparing the actual dynamics with the WW prediction [cf.~dotted lines in Fig.~\ref{Sbfig:NonMark_time_evo_O}(a)] which neglects the coupling to the band edge in this case, as explained in Appendix~\ref{Sbsubsub:retardation}. Finally, we note that in the driven case the continuous stream of emitted magnons tends to damp the transient non-Markovian features appearing as revivals and oscillations. When reaching the steady state, the non-Markovian witnesses tend to zero, giving no longer information. In Sec.~\ref{Sbsub:nonMarksteady}, we identify non-Markovian effects in steady state by quantifying correlations and entanglement. In the case of a bidirectional waveguide, related studies where non-Markovianity is quantified via witnesses or entanglement can be found in \cite{SbTufarelli:2014ku,SbGonzalezBallestero:2013go,SbGulfam:2012ku}.

\subsubsection{Very long time delays, and magnons as hard-core bosons}\label{Sbsub:hardcore}

\begin{figure}[t]
\includegraphics[width=\columnwidth]{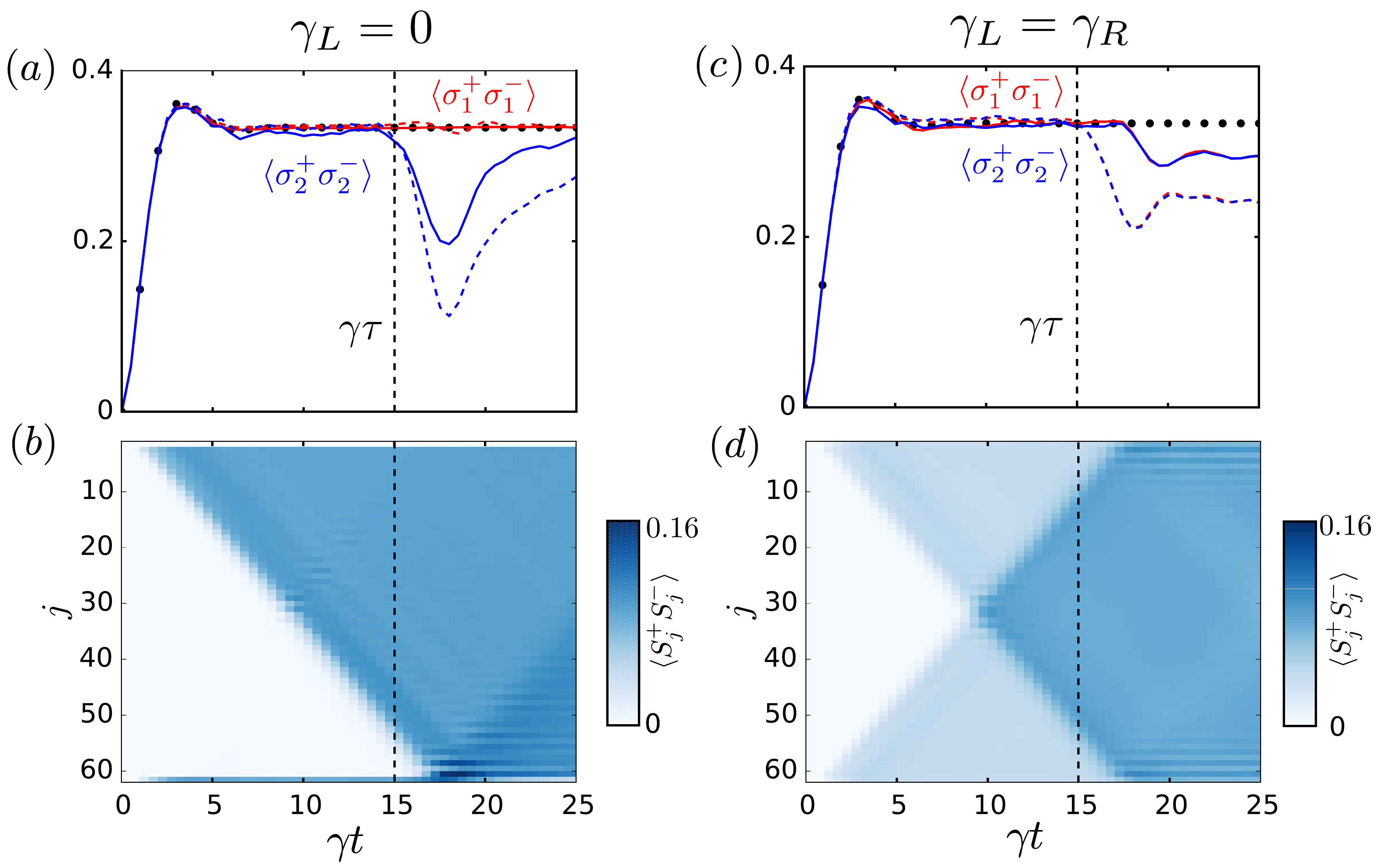}
\caption{Strong hard-core effects in a spin waveguide, with unidirectional (a,b) and bidirectional coupling (c,d). (a) Two strongly driven system spins ($\Omega_\alpha=\gamma$), separated by a distance $d=60a$, and coupled unidirectionally to the waveguide with $\phi=\pi/4$, $\tilde{J}=0.5J$, and $\tilde{\Delta}=0$. (corresponding to $\tau=15/\gamma$). The evolution of the system spin occupations $\langle \sigma^+_\alpha\sigma^-_\alpha\rangle$ are shown in solid lines for a spin waveguide, and in dashed lines for a bosonic waveguide, with red and blue color corresponding to left ($\alpha=1$) and right ($\alpha=2$) system spins, respectively. The black dots represent the Markovian prediction of a single driven system spin ($N_\mathrm{S}=1$), showing that both system spins reach the single-particle quasi-state steady for $t<\tau$. The larger population of the right system spin for $t>\tau$ (blue solid line), compared to the bosonic case (blue dashed line), is due to the saturation of the spin waveguide. (b) For the same situation as in (a), we plot the evolution of the bath spins occupation $\langle S_j^+S_j^- \rangle$, showing the unidirectional emission of magnons for $t<\tau$. At $t>\tau$, the right system spin also emits into left-moving modes as an effect of the high density of the spin waveguide and the hard-core constraint. (c,d) Dynamics of system and bath spins, under the same conditions as in (a,b), but with a bidirectional coupling to the waveguide $\phi=0$. In addition to the saturation effect, the different behavior in the spin (solid) and a bosonic (dashed) case is due to collisions between the magnons. All these simulations were obtained using a MPS representation with a maximum bond dimension of $D=30$ and averaging over $4000$ quantum trajectories, giving a statistical error of less than $0.01$ for the populations shown here. The bosonic waveguide case is obtained when replacing the spin operators $S_j^-$ in Eqs.~(\ref{SbHS})-(\ref{SbHSB}) by bosonic operators $b_j$, and we restrict the number of excitations to a maximum of two per site.
\label{Sbfig:longdelay}}
\end{figure}

In the examples presented so far, the time delay $\tau$ associated with retardation effects is of the order of the time scales $\sim 1/\gamma$, $\sim 1/\Omega$, giving us the possibility to identify deviations from the Markovian regime. For larger time delays, the number of magnons propagating in the waveguide increases, rendering the problem computationally more challenging. Here, we employ MPS methods with the extended Markovian cut (\ref{SbfullMaster}) to keep track of the dynamics in this regime, where the delay $\tau$ represents the largest time scale \footnote{We remark that standard MPS methods for 1D spin chains are efficient to account for the generated system-bath entanglement since all the couplings and Lindblad terms involved in our network model are local.}.

In particular, we investigate the transient dynamics of a network with very {\em long delay lines}, where each system spin evolves first \emph{independently}, reaching a quasi-steady state at $t_\mathrm{qss}\sim 1/\gamma$, before the stream of emitted magnons reaches the other system spin at $\tau\gg t_\mathrm{qss}$. Such a situation, where nodes interact via a time delay, is reminiscent of a `quantum feedback' problem, recently addressed in the photonic context with alternative methods~\cite{SbGrimsmo2015,SbPichler2015b}. This regime of high excitation density also provides evidence for the qualitative differences between spin and boson waveguides. In addition to the non-negligible collisional interactions between magnons, the hard-core constraint can also alter the emission properties of system spins due to a saturation of the spin waveguide. 

To distinguish these two spin waveguide effects, we first consider two driven distant system spins ($d=60a$) with a unidirectional coupling ($\gamma_{\rm L}=0$), such that magnons are emitted only in one direction preventing them from colliding with each other. The corresponding dynamics is shown in Figs.~\ref{Sbfig:longdelay}(a,b), where the parameters are chosen such that the distance corresponds to a delay of $\tau=15/\gamma$. As expected, for times $t< \tau$ the system spins evolve independently and both emit in the same direction. However, at $t>\tau$, when the stream of magnons emitted by the left system spin reaches the right one, the latter also starts to emit in the other direction [cf.~Fig.~\ref{Sbfig:longdelay}(b)]. This effect can be explained by the large density of magnons in the vicinity of the second system spin at $t>\tau$, which alters its emission properties as compared to the case of a spin waveguide in the vacuum state. In particular, if the bath spins neighbouring a system spin are excited, the hard-core constraint blocks the transfer of the excitation from the system spin to these bath spins, changing the flux on the plaquettes shown in Fig.~\ref{Sbfig:setup}(b), and thus altering the directionality of emission. We note that this blocking or saturation of the spin waveguide does not only affect the chirality of the system spins, but also inhibits their total emission into the waveguide. This can be seen in Fig.~\ref{Sbfig:longdelay}(a), where we also show the dynamics of the system spins when they are coupled to a non-interacting bosonic waveguide $S_j^-\to b_j$ (dashed lines). In fact, the population of the right system spin at $t>\tau$ is larger compared to the bosonic case, meaning that the spin waveguide is saturated by the large stream of magnons emitted by the left system spin and passing through its position. We note that these hard-core effects of the bath spins vanish when reducing coupling $\tilde J/J$ or driving $\Omega/\gamma$ since then the magnon density in the waveguide becomes small, and thus behave similar to bosons [cf.~Sec.~\ref{Sbchirality_infinite}].

The case of a bidirectional coupling to the spin waveguide ($\gamma_{\rm L}=\gamma_{\rm R}$) is illustrated in Figs.~\ref{Sbfig:longdelay}(c,d). Here, the emitted magnons propagate in both directions and thus collide in the middle of the waveguide, as shown by the ``propagation cone'' in Fig.~\ref{Sbfig:longdelay}(d). The hard-core nature of the spin excitations leads to a {\em reversal of their phase} in each collision~\cite{SbGorshkov2010}, modifying their subsequent absorption by the nodes. This extra collisional $\pi$ phase, in addition to the mentioned saturation, explains the difference in the system dynamics when the interactions are mediated by bidirectional spin and bosonic waveguides [cf.~Fig.~\ref{Sbfig:longdelay}(c)]. In Sec.~\ref{Sbsec:spintronics} we show that this collision-induced phase-shift can be used to implement a {\em quantum phase gate} between two distant system spins.

To finalize the discussion on the transient non-Markovian dynamics, we comment on the efficiency of our MPS method to solve the extended ME (\ref{SbfullMaster}) for long waveguides. To do so, we consider the same example as in Fig.~\ref{Sbfig:longdelay} and estimate the maximum bond dimension $D$ required to accurately represent every quantum trajectory by a MPS. This is related to the amount of entanglement distributed between different partitions of the network, which must be bounded by $\log_2(D)$ to be efficiently represented~\cite{SbDaley2014}. For each trajectory $m$, which is characterized by a set of random quantum jump events, we thus calculate the entanglement entropy $S(\rho_m)=-{\rm Tr}\lbrace \rho_m{\rm log}_2(\rho_m)\rbrace$ for a partition at the middle of the waveguide, defined by $\rho_m(t)={\rm Tr}_{(j>N_{\rm B}/2)}\lbrace|\Psi_m(t)\rangle\langle\Psi_m(t)|\rbrace$. 

\begin{figure}[t]
\includegraphics[width=\columnwidth]{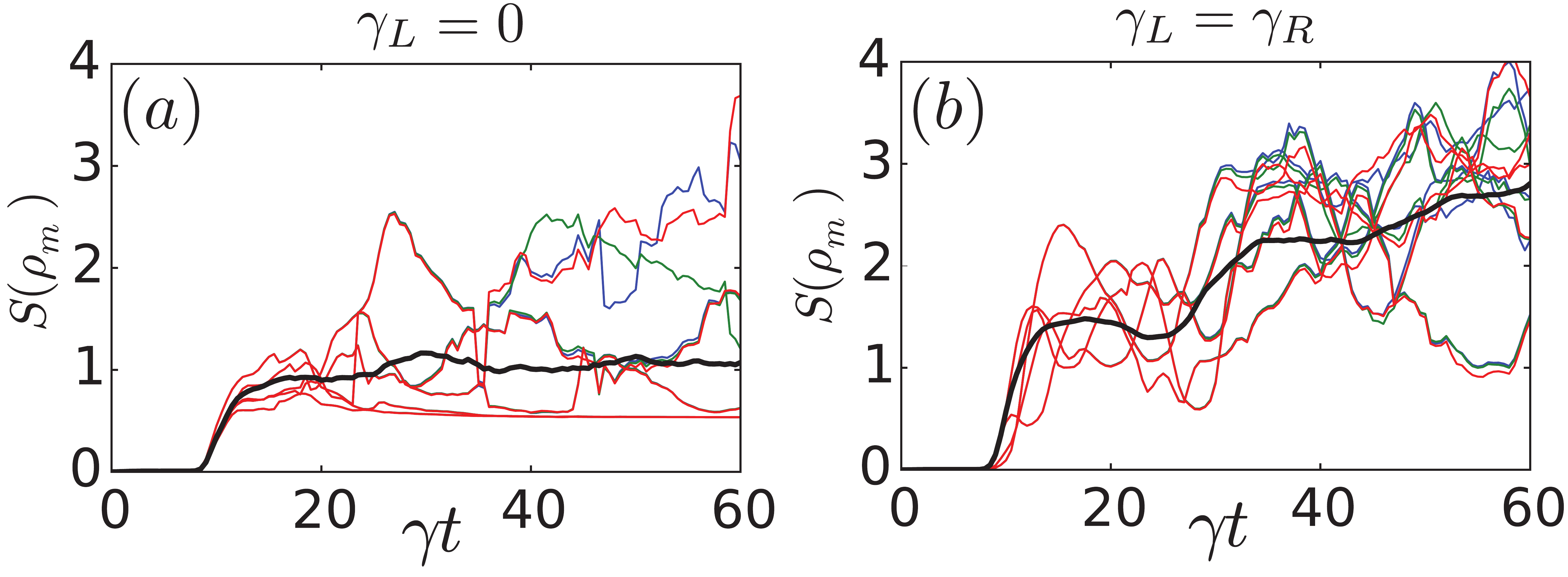}
\caption{Entanglement entropy $S(\rho_m)$ across a bipartite splitting at the middle of the spin chain, for a representative sample of quantum trajectories. We assume the same physical situation and parameters as in Fig.~\ref{Sbfig:longdelay} and check the convergence of the MPS approach by evolving the same quantum trajectory $|\Psi_m\rangle$ for three different maximum bond dimensions $D=30,60,120$ (shown in red, green and blue lines, respectively), and compare their overlaps. For a unidirectional waveguide (a), the average entanglement entropy $\bar{S}$ (black line) remains low during the whole evolution, but at long times some trajectories with large entropy are not converged. In the bidirectional case (b), the progressive entanglement growth also limits the efficiency of the MPS approach at long times. In both cases (a) and (b), the dynamics for $\gamma t<25$ is well described by a bond dimension of $D=30$ [cf.~Fig.~\ref{Sbfig:longdelay}].}
\label{Sbfig:entropytrajectories}
\end{figure}

For a spin waveguide, our results are shown in Fig.~\ref{Sbfig:entropytrajectories} where we compare the entropies $S(\rho_m)$ for a representative sample of trajectories, obtained for three different bond dimensions $D=30,60,120$ (shown as red, green, and blue lines, respectively). The region where curves with different bond dimension $D$, but which correspond to the same trajectory $m$, overlap identifies the time window where the calculations are converged. We see that for times $t\le 25/\gamma$ corresponding to the dynamics shown in Fig.~\ref{Sbfig:longdelay}, all trajectories are well converged with MPSs of maximum bond dimension of $D=30$, in both unidirectional and bidirectional cases. For larger times, however, the entanglement grows further and some trajectories require a larger bond dimension as illustrated by the blue lines not overlapping with green or red ones. Such highly entangled trajectories limit the efficiency of the method, in particular to reach the steady-state. Interestingly, we notice that the quantum trajectories in a bidirectional spin waveguide require a larger average entropy $\bar{S}(t)=\lim_{N\to\infty}\sum_{m=1}^N S(\rho_m)/N$ compared to the unidirectional case [cf.~black lines in Fig.~\ref{Sbfig:entropytrajectories}]. As analyzed in detail in Sec.~\ref{Sbsubsub:nonMark_dimer}, this larger complexity to simulate the evolution with a bidirectional waveguide is related to the possibility of system spins to emit into spin waves propagating in both directions, in contrast to the unidirectional case where they can only be emitted into a single channel.

In Appendix~\ref{Sbapp:MPS_entropy} we extend the present discussion by studying the influence of the retardation time $\tau$ on the entanglement created during the evolution. In addition, we compare the entropies for spin and boson waveguides, and show that in the former case they are smaller due to the absence of hard-core effects.

\subsection{Steady states in the non-Markovian regime}\label{Sbsub:nonMarksteady}

Steady states of driven chiral quantum networks have been previously studied in the Markovian limit \cite{SbStannigel:2012jk,SbRamos2014,SbPichler2015}. One of the most striking predictions is the many-body ``cooling'' of two-level systems into clusterized phases (quantum dimers, tetramers, hexamers, etc.), as the unique steady state of the chiral driven-dissipative dynamics. Below, we will first review the conditions and physical picture behind the formation of such quantum dimers, the simplest clusterized phase, and illustrate their formation and dynamics in the context of the chiral spin waveguides [cf.~Sec.~\ref{Sbsubsub:Mark_dimer}]. The tools developed in the present work will allow us to systematically access the {\em non-Markovian regime of quantum dimer formation} by increasing both the distance between system spins and their coupling to the waveguide. We assess the various signatures of retardation on unidirectional and asymmetric bidirectional (chiral) networks in terms of quantum mutual information between the system spins, two-time correlations, and entanglement entropy [cf.~ Sec.~\ref{Sbsubsub:nonMark_dimer}].  

\subsubsection{Quantum dimer formation in Markovian networks}\label{Sbsubsub:Mark_dimer}

The interplay between driving and dissipation leads the system spins, with dynamics described in the Markovian regime by the ME (\ref{SbchiralMaster}), to a steady state, which is generally mixed. Nevertheless, as pointed out in Refs.~\cite{SbStannigel:2012jk,SbRamos2014,SbPichler2015}, there are special situations when this open system asymptotically decouples from the chiral waveguide, allowing the dissipative formation of pure (and possibly entangled) many-body steady states of the system spins $|\Psi^{\rm ss}\rangle_{\rm S}$, as
\begin{align}
\rho_{\rm S}(t\rightarrow\infty)=|\Psi^{\rm ss}\rangle_{\rm S}\langle\Psi^{\rm ss}|.\label{Sbdissipative_pumping}
\end{align}

\begin{figure}[t]
\includegraphics[width=\columnwidth]{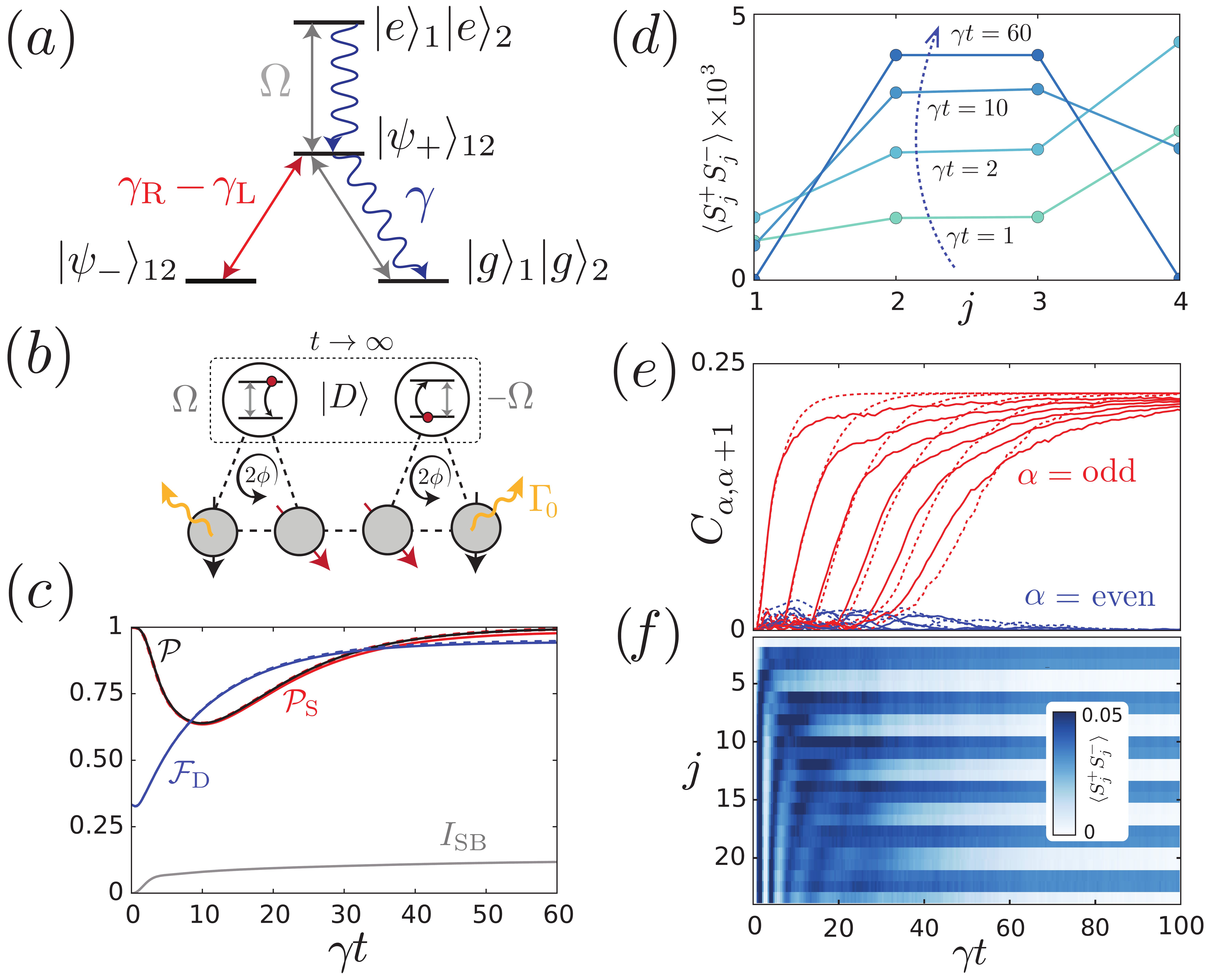}
\caption{Driven-dissipative preparation of pure dimer states, in the Markovian regime (a) Level scheme and couplings, in the Markovian limit, and under the conditions for the dimer formation between two system spins. (b) Minimal configuration for dimer formation, using $N_\mathrm{S}=2$, $N_{\rm B}=4$, $d=2a$, $\tilde{\Delta}=0$, and $\Omega_1=-\Omega_2=\Omega$. (c) Dynamical formation of dimer steady state illustrated by the evolution of the purity $\mathcal{P}_{\rm S}$ of the system density matrix and its overlap with the dimer state $\mathcal{F}_\mathrm{D}$. Both approach unity, in good agreement with the Markovian prediction (dashed lines). The imperfect decoupling between system and waveguide is detected by a non-zero mutual information $I_{\rm SB}$ (grey line). The purity ${\cal P}$ of the whole $2+4$ spin network also approaches unity, signaling the formation of a composite dark state between system and bath spins. Other parameters are $\tilde{J}=0.1J$, $\Omega/\gamma=1$, and $\phi=\pi/10$. (d) For the same situation as in (c), we show snapshots of the bath spin occupations $\langle S_j^+S_j^-\rangle$, evidencing the formation of a constant magnon flux between the system spins, as $\gamma t\to\infty$. (e) Dissipative formation of a dimerized phase of $N_\mathrm{S}=12$ system spins, illustrated by the singlet correlations $C_{\alpha,\alpha+1}$, which build up in pairs, as predicted in the Markovian limit (dashed lines). Other parameters are $\tilde{J}=0.3J$, $\Omega/\gamma=1$, $\tilde{\Delta}=0$, $\phi=\pi/4$, and $N_{\rm B}=24$. (f) For the same situation as in (e), we show the evolution of the bath spin occupations $\langle S_j^+S_j^-\rangle$, showing the decoupling between dimers in steady state, which occurs successively from left to right due to the unidirectional coupling. The dynamics shown in (e,f) was calculated using an MPS representation with a maximum bond dimension of $D=10$ and averaging over $400$ quantum trajectories, giving a statistical error of less than $0.01$ for the occupations shown here.}\label{Sbfig:Marksteadystate}
\end{figure}

The simplest network to illustrate this mechanism is again two driven nodes chirally coupled to the spin waveguide, which are dissipatively purified into an entangled \emph{dimer} steady state [cf.~Sec.~\ref{Sbsec:markov}]. This is achieved in the Markovian limit, and in the specific situation where the effective coherent and dissipative couplings, given in Eqs.~(\ref{SbHS}), (\ref{SbchiralH}) and (\ref{SbglobalJump}), act on the system states as shown in Fig.~\ref{Sbfig:Marksteadystate}(a), where the triplet $(+)$ and the singlet $(-)$ states are defined as 
\begin{align}
|\psi_{\pm}\rangle_{12}=\left(|e\rangle_1 |g\rangle_2\pm e^{i\bar{k}d}|g\rangle_1|e\rangle_2\right)/\sqrt{2}.
\end{align} 
The first condition is to drive the two system spins on resonance ($\tilde{\Delta}=0$) and with proper phases, $\Omega_\alpha=e^{i\alpha\bar{k}d}\Omega$, such that the singlet state decouples from the drive [cf.~Fig.~\ref{Sbfig:Marksteadystate}(a)]. Second, we require a commensurate distance between the two system spins, $\bar{k}d=n\pi$ (with $n$ an integer), such that the singlet $|\psi_{-}\rangle_{12}$, is annihilated by both jump operators $c_{\rm R}=c_{\rm L}=\sigma_2^{-}+(-1)^{n}\sigma_1^{-}$, and thus becomes perfectly subradiant \cite{SbStannigel:2012jk,SbPichler2015}. We notice that in the present spin context, the commensurability condition simply reduces to $d/2a=n$. Finally, we need the decay into the waveguide to be chiral $\gamma_{\rm L}<\gamma_{\rm R}$, so that the effective Hamiltonian $H_{\rm C}$ in Eq.~(\ref{SbchiralH}) couples singlet and triplet, which in combination with $|g\rangle_1|g\rangle_2$ resemble a $\Lambda$-system [cf.~Fig.~\ref{Sbfig:Marksteadystate}(a)]. This admits a unique dark state as a superposition between $|\psi_{-}\rangle_{12}$ and $|g\rangle_1|g\rangle_2$, which decouples from the coherent dynamics due to destructive interference between the drive, decay and interactions. Explicitly, it is given by
\begin{align}
|D\rangle_{12}=\frac{1}{\sqrt{1+|{\cal S}|^2}}\left(|g\rangle_{1}|g\rangle_2+{\cal S}|\psi_{-}\rangle_{12}\right),\label{Sbdimer_state}
\end{align}
where the singlet fraction reads
\begin{align}
{\cal S}=-i\sqrt{2}\left(\frac{\Omega}{\gamma}\right)\frac{(1+\gamma_{\rm L}/\gamma_{\rm R})}{(1-\gamma_{\rm L}/\gamma_{\rm R})}.\label{Sbsinglet_fraction}
\end{align}
In analogy to optical pumping \cite{SbHAPPER:1972fi}, for long times all the population will be pumped into the dimer dark state $|D\rangle_{12}$, being the unique pure steady state of the driven-dissipative dynamics, as in Eq.~(\ref{Sbdissipative_pumping}).

The emergence of this dimerization and the corresponding dynamics of the waveguide excitations can be clearly seen in the full evolution of the network, by including the spin waveguides via the extended Markovian cut (\ref{SbfullMaster}), but choosing parameters so that we remain in the Markov regime. Remarkably, we find that all the features of the chiral ME \eqref{SbchiralMaster} can be recovered in a minimal model consisting of only two spins in the waveguide per spin in the system. This is illustrated in Fig.~\ref{Sbfig:Marksteadystate}(b), where four bath spins are enough to account for the collective reservoir-mediated effects, while a single loss on each end, with rate $\Gamma_0=2J$, perfectly mimics the output of an infinite (non-reflecting) waveguide [cf.~Appendix \ref{Sbapp:FiniteMaster}]. 

For this minimal setup, we plot in Fig.~\ref{Sbfig:Marksteadystate}(c) the dimer fidelity ${\cal F}_{\rm D}(t)={\rm Tr}\lbrace \rho_{\rm S}|D\rangle_{12}\langle D|\rbrace$ and the system purity ${\cal P}_{\rm S}(t)={\rm Tr}\lbrace \rho_{\rm S}^2\rbrace$, as a function of time, with $\rho_{\rm S}(t)={\rm Tr}_{\rm B}(\rho)$ the reduced system state, and $|g\rangle_1|g\rangle_2|0\rangle$ the initial condition. The time evolution from Eq.~(\ref{SbfullMaster}) agrees well with the Markovian prediction, reaching the pure dimer steady state with high fidelity. The direct access to the waveguide allows us to easily quantify system-bath correlations via their quantum mutual information, defined as $I_{\rm SB}=S(\rho_{\rm S})+S(\rho_{\rm B})-S(\rho)$. Here, $\rho_{\rm B}$ denotes the reduced state of the bath spins alone and $S(\rho)=-{\rm Tr}\lbrace \rho\log_2(\rho)\rbrace$ the von Neumann entropy. The corresponding time evolution is shown in Fig.~\ref{Sbfig:Marksteadystate}(c), where the small value of $I_{\rm SB}$ reached in steady state witnesses a slightly imperfect decoupling between system and waveguide. Nevertheless, $I_{\rm SB}$ reduce with coupling $\tilde{J}/J$, as the Markov approximation becomes more valid. Interestingly, the total purity of system and waveguide together, ${\cal P}(t)={\rm Tr}\lbrace \rho^2\rbrace$, reaches a larger steady state value than ${\cal P}_{\rm S}$, meaning that the entire $2+4$ spin network also decouples from the output, and forms a better dark state than when considering the system spins alone.

The interference effect underlying the dimer dark state formation is evident when looking at the waveguide dynamics. In particular, magnons should be stationary exchanged between system spins forming a dimer, without being able to escape from the pair as they are perfectly absorbed. This is clearly seen in Fig.~\ref{Sbfig:Marksteadystate}(d), where a constant occupation in the bath spins $j=2,3$ is dynamically built-up, whereas the other bath spins $j=1,4$, sitting outside the nodes forming the dimer become completely depopulated in steady state, i.e.~the output becomes `dark' and magnons are no longer emitted into the region outside the pair.

Under the same conditions as discussed above, but for an arbitrary even number $N_{\rm S}$ of system spins, the chiral ME (\ref{SbchiralMaster}) predicts the dissipative formation of a large dimerized pure steady state, $|\Psi^{\rm ss}\rangle_{\rm S}=\bigotimes_{\alpha=1}^{N_{\rm S}/2}|D\rangle_{2\alpha-1,2\alpha}$, where each system spin pairs up with one of its neighbors in the dimer state (\ref{Sbdimer_state}), and completely decouples from all the others \cite{SbStannigel:2012jk,SbRamos2014,SbPichler2015}. This is illustrated in Figs.~\ref{Sbfig:Marksteadystate}(e,f), where we calculate the dynamics for $N_\mathrm{S}=12$ driven system spins, coupled to $N_{\rm B}=24$ bath spins with unidirectional interactions ($\gamma_{\rm L}=0)$. Here, the adjacent spin correlations, defined as
\begin{align}
C_{\alpha,\alpha+1}=|\langle \sigma_{\alpha+1}^+\sigma_{\alpha}^-\rangle-\langle\sigma_{\alpha+1}^+\rangle\langle\sigma_\alpha^-\rangle|,\label{SbsingletCorr}
\end{align}
with $\alpha=1,\dots,N_{\rm S}-1$, witness the onset of the dimerized phase in steady state. As shown in Fig.~\ref{Sbfig:Marksteadystate}(e), spin correlations $C_{\alpha,\alpha+1}$ between neighboring system spins forming a dimer (for $\alpha={\rm odd}$), asymptotically approach to the same value given by
\begin{align}
C_{1,2}(t\rightarrow\infty)=\frac{(-1)^{1+d/2a}|{\cal S}|^4}{2(1+|{\cal S}|^2)^2},\label{SbsteadystateCorr}
\end{align}
while the other adjacent correlations (for $\alpha={\rm even}$) vanish in steady state, witnessing the decoupling of system spins not forming the same dimer. Due to the unidirectional coupling, this dimerization occurs successively from left to right, which is clearly seen in the dynamics of $C_{\alpha,\alpha+1}(t)$, but also in the dynamical build-up of the step-like bath occupation between dimers, shown in Fig.~\ref{Sbfig:Marksteadystate}(f). We note that the local entanglement structure of this dimerized phase is efficiently represented by MPSs, which allows us to account for large system sizes within our extended ME framework.

\subsubsection{Non-Markovian effects in the dimer steady state}\label{Sbsubsub:nonMark_dimer}

Including the dynamics of the spin waveguide with our extended Markovian cut allows us to systematically investigate the steady states of chiral networks deep in the non-Markovian regime. We illustrate this for the case of $N_{\rm S}=2$ system spins under the commensurability conditions, which leads to dimer formation in the Markovian limit [cf.~Sec.~\ref{Sbsubsub:Mark_dimer}]. Here, the system spins share correlations in steady state, as discussed in the singlet correlations (\ref{SbsteadystateCorr}) [see also Fig.~\ref{Sbfig:Marksteadystate}(e)]. In the non-Markovian regime, however, we find that the dark state formation is imperfect, leading to an incomplete decoupling of the nodes from the waveguide, and thus to a reduction of the correlations between system spins, or decoherence.
 
We quantify the correlations between the two system spins via the quantum mutual information as,
\begin{align}\label{Sbeq:mutual}
I_{12}=S(\rho_1^{\rm ss})+S(\rho_2^{\rm ss})-S(\rho_{\rm S}^{\rm ss}).
\end{align}
Here $\rho_{\rm S}^{\rm ss}={\rm Tr}_{\rm B}\lbrace\rho(t\rightarrow\infty)\rbrace$ is the reduced density matrix in steady state of both system spins, and $\rho_1^{\rm ss}$, $\rho_2^{\rm ss}$ is the reduced density matrix of node $1$ and $2$, respectively. We are interested in studying the robustness of the dimer correlations in the non-Markovian regime, for which $I_{12}$ is an appealing measure, as it does not assume any specific entanglement structure of the resulting steady state $\rho_{\rm S}^{\rm ss}$. More specifically, it can take values in the range $0\leq I_{12}\leq 2$, where the maximum correlations $I_{12}=2$ are achieved in the case of a maximally entangled state between the two system spins. We note that the non-Markovian witnesses used in Sec.~\ref{Sbsub:transient} do not apply in steady state, as they are based on (transient) time evolution, and thus can not quantify non-Markovianity in this situation. 

\begin{figure}[t]
\includegraphics[width=\columnwidth]{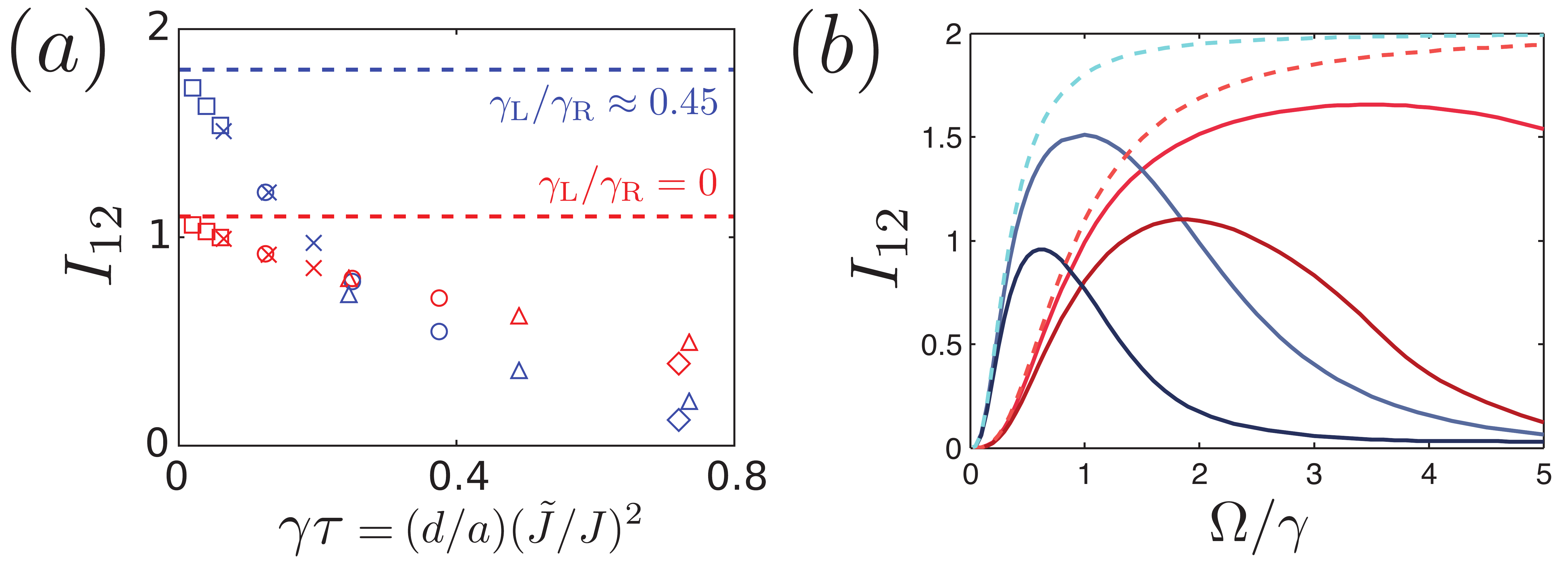}
\caption{Robustness of steady state correlations in the Non-Markovian regime. (a) Quantum mutual information $I_{12}$ between the two system spins in steady state, as a function of $\gamma\tau=(d/a)(\tilde{J}/J)^2$, for fixed ratio $\Omega/\gamma=1$, and two values of chirality: $\phi=\pi/4$ (unidirectional, red symbols), and $\phi=\pi/16$ (asymmetric bidirectional, blue symbols). The various symbols correspond to different combinations of distance and couplings given by $d/a=[2,4,6]$ and $\tilde{J}/J=[0.1,0.18,0.25,0.35,0.6]$, where the different values of $\tilde{J}/J$ are represented by squares, crosses, circles, triangles, and diamonds, with increasing value. The Markovian predictions are represented as dashed lines, showing that the maximal correlations are obtained in this limit ($\tau=0^+$). The linear scaling of $I_{12}$ with $\gamma\tau$ in the region $\gamma\tau\ll1$ identifies retardation effects as the main source of deviation from the Markovian prediction. (b) Mutual information $I_{12}$ as a function of $\Omega/\gamma$, for $\phi=\pi/4$ (red) and $\phi=\pi/16$ (blue) cases, and two values of retardation $\gamma\tau=[0.065,0.245]$ (increasing for darker color), corresponding to $\tilde{J}/J=[0.18,0.35]$ and $d/a=2$. There is an optimal $\Omega/\gamma$, at which the correlations are maximal, which reduces with $\gamma\tau$ and $\gamma_{\rm L}/\gamma_{\rm R}$. The dashed lines show the Markovian predictions for each chirality, which saturate to the maximum value $I_{12}\rightarrow 2$ for $\Omega/\gamma\to\infty$, as the dimer state coincides with the singlet.}\label{Sbfig:nonMark2_steady_state}
\end{figure}

Non-Markovian effects become visible first of all by increasing the distance between the two system spins $d$, and thus increasing the time $\tau$ that excitations need to propagate from one node to the other. Additionally, a larger coupling to the waveguide $\tilde{J}$ enhances their emission rate $\gamma$. Nevertheless, both have the same effect to first order, since the relevant physical parameter is the product $\gamma\tau=(d/a)(\tilde J/J)^2$, i.e.~the number of excitations travelling between the nodes along the waveguide. For $\gamma\tau\ll 1$ and $d/a=2n$ with $n$ an integer, the local coupling of each node to the waveguide is necessarily weak, $(\tilde J/J)^2\ll 1$, and to leading order all non-Markovian deviations stem from time delays in the interactions. This can be seen in Fig.~\ref{Sbfig:nonMark2_steady_state}(a) where we plot the steady state mutual information $I_{12}$, for various combinations of parameters. Specifically, the correlations present in the dimer for $\gamma\tau\rightarrow 0^+$ decrease linearly with $\gamma\tau$ as the two system spins get entangled with an increasing number of bath spins. For larger couplings $\gamma\sim 2J$, dispersion and band-edge effects are also important, such that $I_{12}$ does no longer scale as $\gamma\tau$. This behavior is presented in Fig.~\ref{Sbfig:nonMark2_steady_state}(a) for two values of chirality, $\gamma_{\rm L}/\gamma_{\rm R}=0$ (red line) and $\gamma_{\rm L}/\gamma_{\rm R}\approx 0.45$ (blue line), showing that correlations in the unidirectional limit are more robust against retardation.

We note that -- in contrast to the Markovian prediction (\ref{SbsteadystateCorr}) -- it is not possible to achieve the maximal steady state correlations between the two system spins by simply increasing the driving strength $\Omega/\gamma$ [cf.~Fig.~\ref{Sbfig:nonMark2_steady_state}(b)]. While the chiral ME for Markovian networks (\ref{SbchiralMaster}) predicts a monotonic increase of $I_{12}$ due to a larger singlet fraction (\ref{Sbsinglet_fraction}), when including a finite time delay $\gamma\tau$ the correlations between the system spins show a qualitatively different behavior. The two nodes decorrelate at large driving and there is an optimal driving strength at which the largest correlations are achieved. The correlations at the optimal driving are, however, always smaller that the maximal value of $I_{12}=2$ for the perfect singlet and reduce with increasing retardation $\gamma\tau$ and $\gamma_{\rm L}/\gamma_{\rm R}$, as can be seen in Fig.~\ref{Sbfig:nonMark2_steady_state}(b).

\begin{figure}[ht!]
\includegraphics[width=\columnwidth]{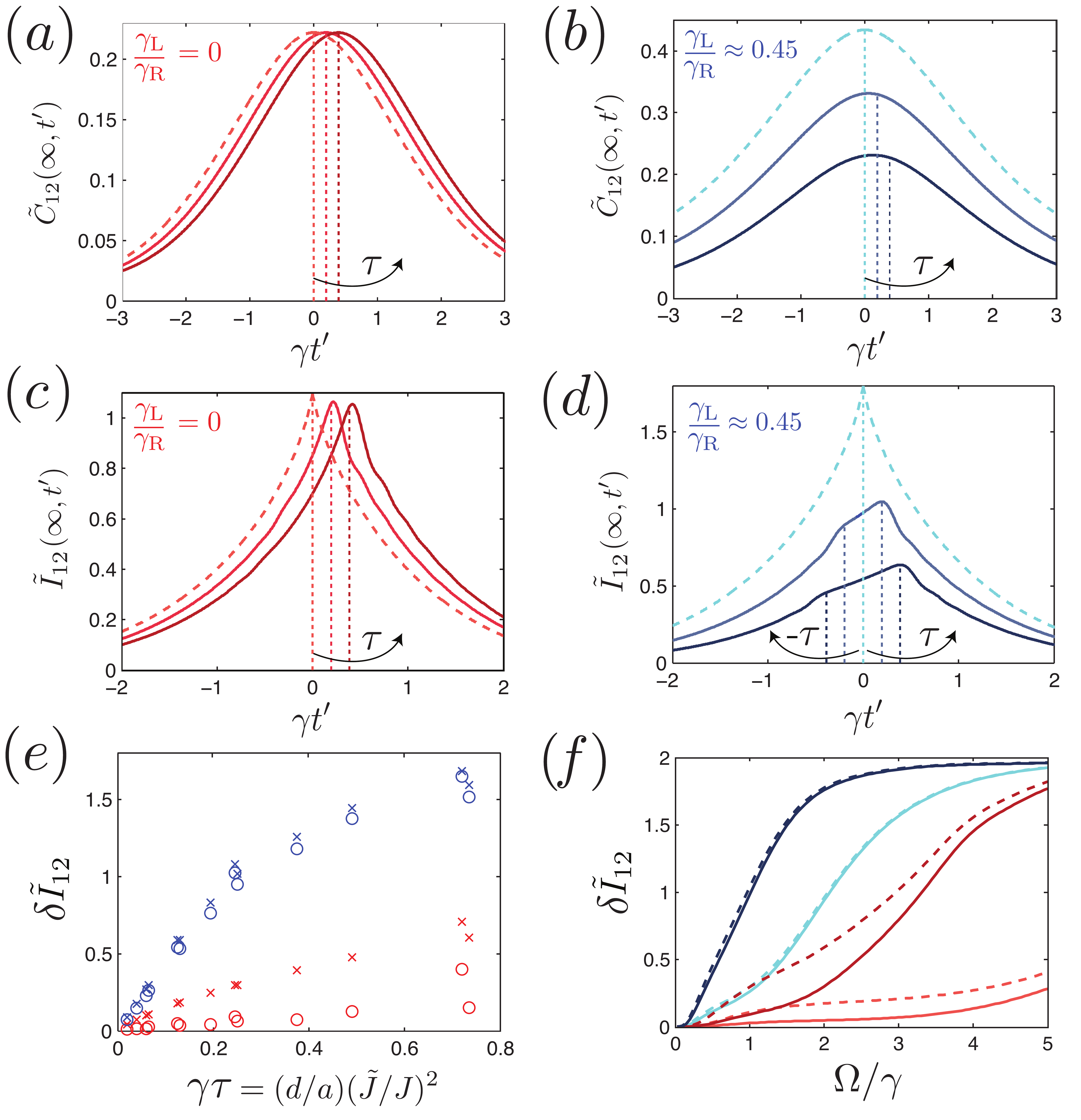}
\caption{Non-Markovian effects in the dimer steady state via two-time correlations. (a,b) Two-time singlet correlation function $\tilde{C}_{12}(\infty,t')$ in steady state, for different delay times $\tau\approx 0.19/\gamma$ and $\tau\approx 0.39/\gamma$ (increasing for darker color), in the unidirectional $\phi=\pi/4$ (a), and in an asymmetric bidirectional case $\phi=\pi/16$ (b). (a) For $\gamma_{\rm L}=0$, the correlations are rigidly displaced by $t'=\tau$ as compared to the Markovian prediction (dashed), showing that the effects of retardation can be compensated by a simple time shift of $\tau$. (b) For $\gamma_{\rm L}/\gamma_{\rm R}\approx 0.45$, there is a strong overall reduction of the two-time correlations with $\tau$, relative to the Markovian prediction (dashed). Other parameters are $d/a=[6,12]$, $\tilde{J}/J=0.18$, $\Omega/\gamma=1$, and $\tilde{\Delta}=0$. (c,d) `Two-time' mutual information $I_{12}(\infty,t')$ in steady state, for the same parameters as in (a,b). (c) In addition to the expected rigid shift by $\tau$, we observe a small decrease in the maximum of $I_{12}(\infty,t')$ due to residual dispersion effects. (d) Besides showing the overall reduction of correlations with $\tau$, $I_{12}(\infty,t')$ is also sensitive to two peaks at $t'=\pm\tau$, showing that correlations can be restored by a time shift in either direction, the maximum of the two obtained in the direction of chirality. (e,f) Deviation of the maximal `two-time' mutual information $\delta\tilde{I}_{12}=I_{12}^{\rm mark}-\tilde{I}_{12}(\infty,t'_{\rm max})$ and of the equal time one $\delta I_{12}=I_{12}^{\rm mark}-I_{12}$, with respect the Markovian prediction $I_{12}^{\rm mark}$ obtained for $\tau=0^+$. (e) We plot $\delta\tilde{I}_{12}$ (circles) and $\delta I_{12}$ (crosses), as a function of $\gamma\tau=(d/a)(\tilde{J}/J)^2$, for the same paramaters as in Fig.~\ref{Sbfig:nonMark2_steady_state}(a). In particular, red symbols correspond to $\gamma_{\rm L}=0$, while blue symbols to $\gamma_{\rm L}/\gamma_{\rm R}\approx 0.45$. For $\gamma_{\rm L}=0$, $\tilde{I}_{12}(\infty,t'_{\rm max})$ allows to extract correlations much closer to the Markovian prediction than $I_{12}$, whereas for $\gamma_{\rm L}/\gamma_{\rm R}\approx 0.45$ the correlations are only slightly increased. (f) The same conclusions from (e) can be drawn when plotting $\delta\tilde{I}_{12}$ (solid) and $\delta I_{12}$ (dashed), as a function of driving $\Omega/\gamma$, for the same paramaters and color code as in Fig.~\ref{Sbfig:nonMark2_steady_state}(b).\label{Sbfig:nonMark2_steady_stateTwoTime}}
\end{figure}

In the {\em unidirectional} (cascaded) case ($\gamma_{\rm R}>\gamma_{\rm L}=0$) and weak coupling limit ($\gamma\ll 2J$) the effect of retardation is particularly simple to understand, as waveguide excitations emitted by the first node propagate to the second node, but not vice versa. In the absence of this back-action, we expect that the {\em time delay} $\tau=d/|\bar{v}|$ will {\em shift in time the effective dynamics of the second system spin}, when compared to the Markovian prediction (corresponding to $\tau=0^+$). In the theory of cascaded quantum systems according to Refs.~\cite{Sbgardiner2015,SbGardiner1993,SbCarmichael1993} the effect of an arbitrary time delay $\tau$ is accounted for by interpreting `time' as the `retarded time' for the second node as $\sigma^{\pm}_2(t+\tau)\rightarrow\sigma^{\pm}_2(t)$. Therefore, in the case of an unidirectional coupling it is possible to recover the lost equal time dimer correlations [shown in Fig.~\ref{Sbfig:nonMark2_steady_state}(a)] in {\em two-time correlations} with finite time delay $\tau$. This is clearly seen in Fig.~\ref{Sbfig:nonMark2_steady_stateTwoTime}(a), where in analogy to the {\em equal time correlation}  (\ref{SbsingletCorr}) we plot the {\em two-time correlation function} 
\begin{equation}
\tilde{C}_{12}(t,t')=|\langle \sigma_2^+(t+t')\sigma_1^-(t)\rangle-\langle\sigma_2^+(t+t')\rangle\langle\sigma_1^-(t)\rangle|,\label{SbtwotimeCorrSin}
\end{equation}
as a measure of the delayed singlet correlations of the two system spins, which is evaluated in steady state $t\rightarrow\infty$ for various delay times $\tau$. In the Markovian limit ($\tau=0^+$, dashed line) we observe the expected maximum of correlations at $t'=0$, and the vanishing of two-time correlations for $|t'|\gg 1/\gamma$. A finite time delay indeed leads to a rigid displacement by $\tau$ of the correlations predicted by the Markovian theory of cascaded systems: the maximum value is now obtained at the delayed time $t'=\tau$.

In contrast, for an {\em asymmetric bidirectional} case correlations between nodes are established by excitations propagating in both directions ($\gamma_{\rm R}>\gamma_{\rm L}\neq0$). Figure~\ref{Sbfig:nonMark2_steady_stateTwoTime}(b) shows the corresponding two-time correlation function (\ref{SbtwotimeCorrSin}) in steady state, for various time delays $\tau$ as above. Although the maximum correlations appear at a slightly delayed time $t^\prime \lesssim \tau$, there is a strong overall reduction of correlations with increasing $\tau$ relative to the Markovian limit $\tau=0^+$.
 
We note that the mutual information (\ref{Sbeq:mutual}) discussed above measures quantum correlation between the two system spins at {\em equal times}. In light of our discussion of two-time correlations (\ref{SbtwotimeCorrSin}) with maximum quantum correlation at a finite time difference related to $\tau$, one may wonder if one can introduce an analogous measure based on mutual information for two-time correlations. This can be done adopting and generalizing arguments by Carmichael \cite{SbCarmichael1993}: consider the whole network of node $1$, node $2$, and spin waveguide described at time $t$ by the density operator $\rho(t)$, as solution of the extended ME (\ref{SbfullMaster}). We remove node $1$ at time $t$ from the dynamics of the total system, and we define a density operator for the following evolution as 
\begin{equation}
\tilde{\rho}(t,t') = e^{{\cal L}_2t'}\rho(t)\quad(t'>0),\label{SbshiftedDensityMat}
\end{equation}
as a function of time $t'>0$.  Here ${\cal L}_2$ is the truncated Liouvillian obeying an extended master equation (\ref{SbfullMaster}) with the first node removed from the dynamics. Physically, $\tilde{\rho}(t,t')$ describes a situation where node $2$ will for the time interval $t'<\tau$ still `see' all the waveguide excitations, which were emitted by node $1$ at times earlier than $t$ (i.e.~before the node $1$ was decoupled), and which continue propagating towards node $2$. We emphasize that $\tilde{\rho}(t,t')$ is a proper density operator, and we take it as an operational definition of a `two-time quantum state'. In addition, it can be naturally extended to negative times $t'<0$, by interchanging the roles of nodes $1$ and $2$ in Eq.~(\ref{SbshiftedDensityMat}).

The idea is now to define a `two-time' quantum mutual information, in analogy to $I_{12}$, as
\begin{align}\label{Sbeq:mutualTwoTime}
\tilde{I}_{12}(t,t')=S(\tilde{\rho}_1(t,t'))+S(\tilde{\rho}_2(t,t'))-S(\tilde{\rho}_{\rm S}(t,t')),
\end{align}
where $\tilde{\rho}_{\rm S}(t,t')={\rm Tr}_{\rm B}\lbrace\tilde{\rho}(t,t')\rbrace$ is the `two-time' reduced density matrix of both system spins and $\tilde{\rho}_1(t,t')$, $\tilde{\rho}_2(t,t')$ the reduced states of system spins $1$ and $2$, respectively. In the stationary limit $t\rightarrow \infty$, and for various time delays $\tau$, we discuss $\tilde{I}_{12}(t,t')$ as a function of $t'$, in analogy to the two-time correlation function $\tilde{C}_{12}(t,t')$ above. The corresponding results can be found in Figs.~\ref{Sbfig:nonMark2_steady_stateTwoTime}(c,d), which look qualitatively similar to the ones in Figs.~\ref{Sbfig:nonMark2_steady_stateTwoTime}(a,b). Nevertheless, as $\tilde{I}_{12}$ does not assume any specific form of two-time correlations, it allows us to distinguish additional phenomena not present in $\tilde{C}_{12}$. In the unidirectional case [cf.~Fig.~\ref{Sbfig:nonMark2_steady_stateTwoTime}(c)], besides the rigid displacement by $\tau$ of the correlations predicted by the Markovian theory ($\tau=0^+$), we observe a small decrease of the shifted maxima due to higher order effects stemming from the non-linear Bloch band dispersion relation. In the asymmetric bidirectional case [cf.~Fig.~\ref{Sbfig:nonMark2_steady_stateTwoTime}(d)], we clearly identify two peaks of the correlations at $t'=\pm\tau$, in addition to the overall decrease of correlations with $\tau$ relative to the Markovian prediction. Since the chirality is chosen with preference to the right ($\gamma_{\rm R}>\gamma_{\rm L}\neq0$), more waveguide excitations are emitted from node $1$ to $2$, and thus the maximal correlations are obtained at the positive delayed time $t'=\tau$. However, the smaller peak at $t'=-\tau$ evidences that the same time shift argument independently applies to the fewer left-moving magnons emitted from node $2$ to $1$. In essence, as the `two-time' steady state $\tilde{\rho}_{\rm S}(t\rightarrow\infty,t')$ assumes a privileged delay direction, it can only fully recover the dimer correlations in the unidirectional limit (up to dispersion effects). This is also visible in Figs.~\ref{Sbfig:nonMark2_steady_stateTwoTime}(e,f), where we compare $I_{12}$ with the corresponding `two-time' mutual information $\tilde{I}_{12}(\infty,t'_{\rm max})$ in steady state, for the same parameters as in Figs.~\ref{Sbfig:nonMark2_steady_state}(a,b), and optimizing the delay time $t'_{\rm max}\approx \tau$ such that the two-time correlations between the nodes are maximal. While in the unidirectional case this allows us to extract two-time correlations that are much closer to the Markovian value than $I_{12}$ (crosses and dashed lines), in the more bidirectional case the delayed correlations increase only by a small amount.

\begin{figure}[t]
\includegraphics[width=\columnwidth]{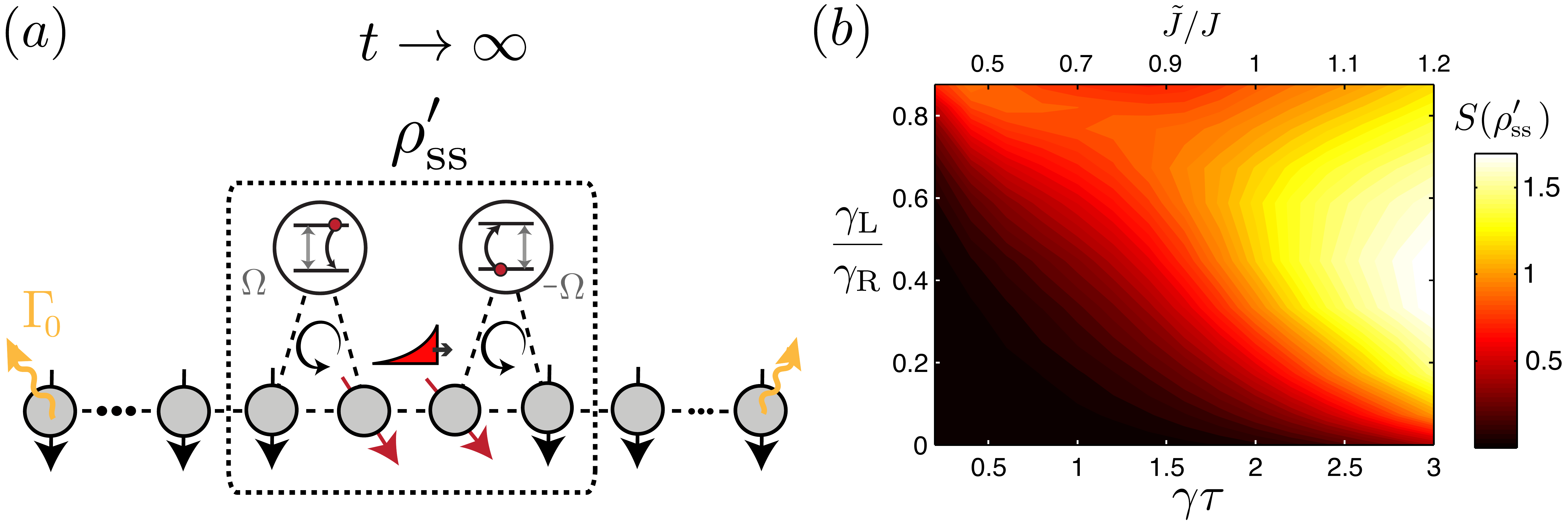}
\caption{Steady state entropy between system and waveguide (a) Relevant part of the network composed by the system spins and only the bath spins connecting them (enclosed by dashed lines). (b) Steady state entropy $S(\rho'_{\rm ss})$ of the subsystem indicated in (a), as a function of chirality $\gamma_{\rm L}/\gamma_{\rm R}$ and retardation $\gamma\tau$. The vanishing entropy in the limit of weak coupling ($\tilde{J}/J\ll 1$) and unidirectional interactions ($\gamma_{\rm L}/\gamma_{\rm R}\to 0$), signals the formation of a composite dark state (with no output) despite the finite time delay $\tau$. At larger coupling ($\tilde{J}/J\gtrsim 1$), the composite dark state is strongly degraded due to dispersion effects, which is evidenced by an increase of the entropy. Calculations are done for $N_{\rm B}=6$, but $S(\rho'_{\rm ss})$ depends insignificantly on the waveguide size. Other parameters are $d/a=2$, $\Omega=\gamma/4$, and $\tilde{\Delta}=0$.}\label{Sbfig:entropy_strongcoupling}
\end{figure}

In the unidirectional limit, the fact that time delays can be absorbed allows us to understand also the properties of the entire network for $\tau>0$. Whereas the Markovian master equation ($\tau=0^+$) predicts that the system spins decouple from the waveguide and form a pure dark state, for a finite $\tau$ instead it is actually the combined system of the nodes together with the stream of waveguide excitations that form a composite dark state, which disentangles from the output. This is shown in Fig.~\ref{Sbfig:entropy_strongcoupling}(b) where we plot the steady state entropy $S(\rho'_{\rm ss})$ of the network consisting of the nodes and the relevant part of the spin waveguide that connects them, as depicted in Fig.~\ref{Sbfig:entropy_strongcoupling}(a). In particular, the formation of the global dark state is signaled by a vanishing entropy in the unidirectional ($\gamma_{\rm L}/\gamma_{\rm R}\to 0$) and weak coupling region ($\tilde{J}/J\ll 1$). In the bidirectional limit ($\gamma_{\rm L}/\gamma_{\rm R}\to 1$), however, it is not possible to reduce the dynamics to a Markovian one by a simple time shift, and the entire network does not form a dark state when including a finite time delay $\gamma\tau$. Correspondingly, the entropy in Fig.~\ref{Sbfig:entropy_strongcoupling}(a) increases in the bidirectional region. Note that at larger coupling ($\tilde{J}/J\gtrsim 1$), dispersion effects and band-edge effects strongly degrade the dark state formation even in the unidirectional limit, such that the entropy becomes less sensitive to chirality and increases with coupling due to the larger amount of magnons leaving the network.

\section{Quantum spintronic circuits for quantum information applications}\label{Sbsec:spintronics}

From a quantum information perspective, the chiral spin network model proposed in this work provides a natural framework to physically implement, model, and design complex quantum `spin circuits'. Here the spin chains act as chiral quantum channels to realize quantum communication between the nodes of the quantum network, represented by the system spins (as qubits). We note that the theoretical tools of Sec.~\ref{Sbsec:model} to model the quantum network dynamics give the possibility to: (i) systematically go beyond the Born-Markov approximation, allowing in particular to calculate and thus visualize the propagation of the multiple excitations in the spin channels; (ii) to account for effects related to the dispersive nature of a structured reservoir, and imperfections, e.g.~due to the presence of disorder in a spin chain as relevant in solid state realizations;  and (iii) to realize quantum operations exploiting the chiral coupling of nodes to the spin channel in combination with the `hard-core boson' nature of spin excitations, in contrast to the quantum optical realizations with (non-interacting) photons as `flying qubits' in photonic waveguides. 

In this section, we present three illustrative examples of {\em basic building blocks} of chiral spin quantum circuits and quantum protocols. In Sec.~\ref{Sbsub:transfer}, we study state transfer between distant qubits via the spin chain with shaped time-symmetric wave-packets. In Sec.~\ref{Sbsub:timereversal}, we describe a `quantum box', which can be inserted in the spin channel, to `time-reverse' the spin wave-packet propagating in the channel. In particular, this allows for the realization of a state transfer protocol that is resilient to dispersive effects. Finally, in Sec.~\ref{Sbsub:gate}, we exploit the `hard-core' nature of the spin waveguide excitations to realize an entangling quantum gate between two distant qubits. In all these examples, we consider the absence of a driving field $\Omega_{\alpha}=0$, so that the total number of excitations is conserved during the evolution.

\subsection{Quantum state transfer via a spin chain}\label{Sbsub:transfer}

\begin{figure}[t]
\includegraphics[width=0.5\textwidth]{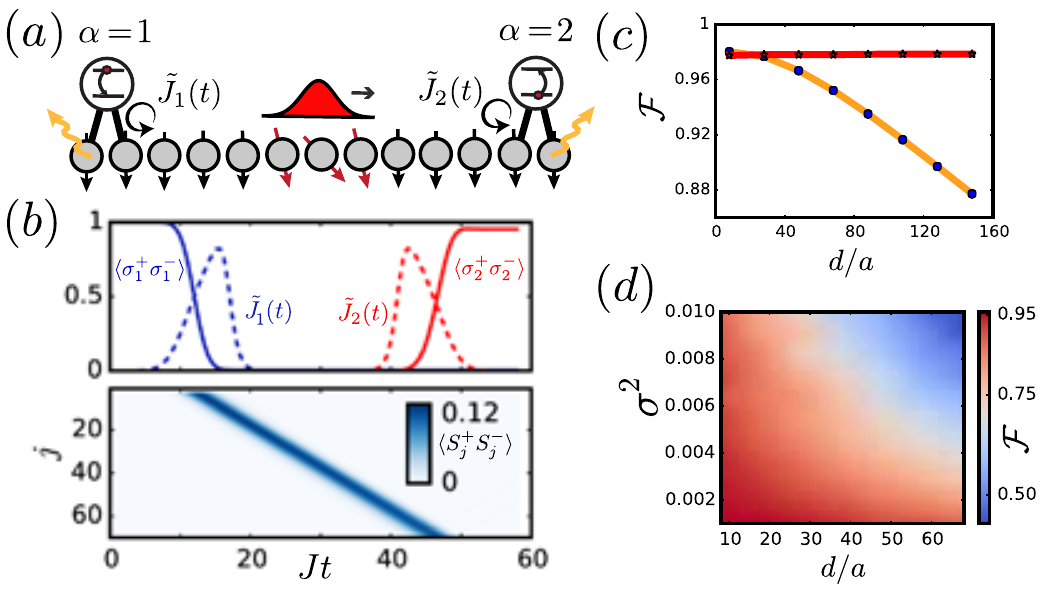}
\caption{State transfer via chiral spin chain. (a) Schematic representation of the quantum state transfer with chiral coupling to the spin chain. (b) Numerical simulation of state-transfer between two system spins (qubits), separated by $d=68a$. The qubit populations are represented in the upper panel (solid lines), together with the pulse shaping of the coupling strengths $\tilde{J}_\alpha(t)$ to achieve a symmetric spin wave-packet (dashed lines). The lower panel shows the propagation of the Gaussian wave-packet in the waveguide. (c) Fidelity of the transfer $\mathcal{F}$ as a function of distance $d$ for a cosine $\omega(k)=-2J \cos(ka)$ (yellow line) and an exactly linear dispersion relation $\omega(k)=2Ja|k|$ (red line). (d) Robustess of the state strasfer protocol in the presence of disorder in the nearest-neighbor hopping of the spin chain $J_j$. We plot the state transfer fidelity $\mathcal{F}$ as a function of $d$ and the variance $\sigma^2$ of the random distribution, showing the destructive impact of disorder at large spatial separations between the qubits.\label{Sbfig:spintronics1} }
\end{figure}

As our first example, we consider a protocol for quantum state transfer between two distant system spins (qubits) via a spin channel with chiral coupling, as illustrated in Fig.~\ref{Sbfig:spintronics1}(a). The goal is to achieve transfer of the state of the first qubit to the second distant qubit via the spin chain, $|\psi\rangle_1|g \rangle_2 \to e^{i\Phi} |g\rangle_1|\psi\rangle_2$,
where $|\psi\rangle_\alpha = c_{g} |g\rangle_\alpha+ c_{e} |e\rangle_\alpha$ refers to an arbitrary state of the qubit $\alpha$, and $\Phi$ accounts for a phase accumulated during propagation. 

We are interested in a spin analogue of the transfer protocol developed in Refs.~\cite{SbCirac1997,SbStannigel2011} for photonic channels. For weak coupling of the qubit to the spin chain, we can approximate the dispersion relation as linear $\omega_k\approx (k-\bar k)|\bar{v}|$ and show analytically~\cite{SbCirac1997,SbStannigel2011} that the protocol can be realized by unidirectionally emitting a {\em symmetric} wave-packet from the first qubit via a time-dependent modulation of the coupling $\tilde{J}_1(t)$, which propagates in the waveguide towards the second qubit. The wave-packet can then be perfectly reabsorbed by the second qubit via a time-reversed coupling $\tilde{J}_2(t)=\tilde{J}_1(\tau-t)$, where $\tau=d/|\bar v|$ represents the time delay [cf  Fig.~\ref{Sbfig:spintronics1} (a)] ~\footnote{Note that in contrast to the model presented in Sec.~\ref{Sbsec:model}, we consider here a situation, where the coupling $\tilde{J}_\alpha(t)$ of the qubit $\alpha$ to the spin chain is not homogeneous.}. In this way, we mimic the time-reversal of the initial decay process~\cite{SbCirac1997}.

Figure~\ref{Sbfig:spintronics1}(b) shows a numerical example illustrating the emission, propagation, and absorption of the wave-packet mediating the state transfer. We emphasize that due to the chiral coupling of the qubits ($\phi=\pi/4$) to the spin waveguide, the two qubits are coupled exclusively to the right moving mode, and we have chosen the separation between the qubits as a large distance $d=68a$. In Fig.~\ref{Sbfig:spintronics1}, the first qubit is assumed to be initially prepared in the excited state $|\psi\rangle_1= |e\rangle_1$, and emits a Gaussian wave-packet [see Appendix.~\ref{Sbapp:gaussianpulse} for details], which propagates in the spin chain during a time $\tau$ before being reabsorbed by the second qubit.

Compared to the ideal situation of a linear dispersion assumed in Refs.~\cite{SbCirac1997,SbStannigel2011}, where the only source of imperfection is due to the finite duration of the coupling pulses $J_\alpha(t)$, here dispersive effects arising from the non-linear dispersion relation of the waveguide degrade the efficiency of the protocol. This is illustrated in Fig.~\ref{Sbfig:spintronics1}(c), where we plot as a function of $d$ the state-transfer fidelity ${\cal F}$, defined by the population of the second qubit at the end of the protocol (yellow curve). Moreover, we remark that when adding suitable long-range couplings to make the dispersion relation exactly linear [see Appendix.~\ref{Sbapp:triangulardispersion}], the resulting fidelity becomes independent of $d$ [cf.~red curve in Fig.~\ref{Sbfig:spintronics1}(c)].

Finally, we estimate the effect of disorder in the spin chain for the quantum transfer protocol. This is motivated by possible implementations of a chiral spin waveguide with solid-state systems~\cite{SbYao2013}, e.g.~spin chains connecting NV centers as qubits. We consider the case of disorder corresponding to a random Gaussian distribution of the nearest-neighbor spin couplings $J_j$ of variance $\sigma^2$. We show in Fig.~\ref{Sbfig:spintronics1}(d), the averaged fidelity over $250$ random distributions, as a function of the distance $d$ and the variance $\sigma^2$. We consider here the case of an exactly linear dispersion relation [cf. Appendix~\ref{Sbapp:triangulardispersion}] allowing us to quantify the parameter regime where the protocol is not affected by disorder~\footnote{We also considered the case of `diagonal disorder' corresponding to a random distribution of the detunings $\Delta_\alpha$, for which we obtained similar results}. Other imperfections in a solid-state context are due to the finite temperature of the spin chain, an effect we will discuss elsewhere.

\subsection{Time-reversing the spin wave-packet and state trasfer resilient to dispersion effects}\label{Sbsub:timereversal}

\begin{figure}[t]
\includegraphics[width=0.5\textwidth]{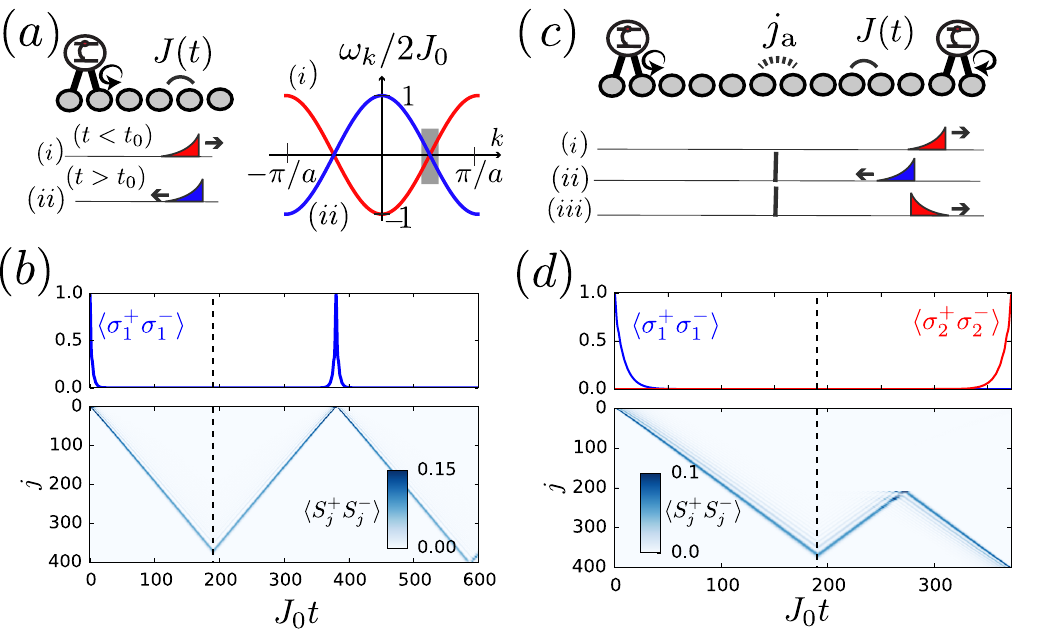}
\caption{Time-reversal and state transfer resilient to dispersion. (a) Schematic illustration of time-reversal of a wave-packet. At $t=t_0$, the sign of the hopping $J(t)$ is reversed, which reverses the sign of the dispersion relation (right) and the direction of propagation of the wave-packet (left). (b) Application of the time-reversal protocol to the spontaneous emission of a system spin into the spin waveguide. Lower panel: The excitation in the waveguide clearly shows the inversion of propagation direction at $t=t_0$, without a change of the wave packet shape. Upper panel: As the system spin population shows, the wave packet is perfectly reabsorbed at $t=2t_0$. Parameters are $\tilde{J}=0.4J_0$, $J_0t_0=190$ and $d=398a$. (c) Application of the time reversal to a state-transfer protocol that is insensitive to dispersive effects. At time $t=t_0$, the emitted wave packet is time-reversed and then reflected due to the presence of a broken link $j_a,j_a+1$. (d) Numerical simulation of the protocol with $\tilde{J}=0.25 J_0$, $J_0t_0=190$ and $d=398a$. Upper panel: Population of the system spins. Lower panel: Occupation of the bath spins in the waveguide, showing the trajectory of the time-reversed wave-packet.\label{Sbfig:timereversal}}
\end{figure}

Instead of shaping a {\em time-symmetric} spin wave-packet via $\tilde{J}_{1}(t)$ as in Sec.~\ref{Sbsub:transfer}, we discuss here a {\em time-reversal} of a spin wave-packet, which consists in reversing the direction of propagation of a wave-packet without altering its shape. The scheme is the spin analogue of a {\em phase-conjugate mirror} in photonics~\cite{SbYanik2004,SbWang2012}, which in the present spin context can be simply achieved by dynamically reversing the sign of the spin chain hopping $J(t)$, during the propagation of the spin wave-packet. Formally, this process reverses the direction of time, as it reverses the spectrum $\omega_k\to -\omega_k$ of the bath Hamiltonian $H_\mathrm{B}$. It allows for a perfect state transfer protocol, which is applicable to arbitrary shapes of wave-packets, and in particular for a wave-packet obtained from an exponential decay of the first qubit, as illustrated in Fig.~\ref{Sbfig:timereversal}(a). Here, for $t<t_0$, the value of the hopping is fixed to $J(t)=J_0$, corresponding to the dispersion relation shown by the red line in Fig.~\ref{Sbfig:timereversal}(a). Consequently, a wave-packet whose momentum distribution is initially centered at $k=\bar k>0$, propagates to the right due to a positive group velocity $(\partial\omega_k/\partial k)(\bar{k})>0$. For $t\geq t_0$ the dispersion relation is reversed by setting $J(t)=-J_0$ (blue line), such that the wave-packet propagates in the opposite direction $(\partial\omega_k/\partial k)(\bar{k})<0$, time-reversing its dynamics.

Figure~\ref{Sbfig:timereversal}(b) shows the most basic example of time-reversal: at time $t=0$, a system spin emits into the right-moving mode via a fixed coupling $\tilde{J}$. The wave-packet propagates in the waveguide until $t_0=190 J_0^{-1}$, when the dispersion relation is reversed. Consequently, the wave-packet comes back to the qubit and is then perfectly reabsorbed, since whole dynamics is completely time-reversed. Remarkably, also the dispersive effects arising from the non-linear dispersion relation are time-reversed and thus compensated.

As an application of this time-reversal `gadget', we now present a state-transfer protocol that is robust against dispersion. The different steps of the protocol are illustrated in Fig.~\ref{Sbfig:timereversal}(c). The first system spin, initially excited, emits a wave-packet to the right. At time $t=t_0$, the dispersion relation is inverted so that the wave-packet then propagates in the left direction. In order to make it reach the second qubit, an additional reflection is required. To do so, we remove the hopping term of a specific link $j_\mathrm{a},j_\mathrm{a}+1$ during the second part of the process at some time $t>t_0$, thus cutting the spin chain into two distinct pieces [cf.~Fig.~\ref{Sbfig:timereversal}(c)]. Consequently, the wave-packet reaching the site $j_\mathrm{a}+1$ is reflected and finally moves in the right direction before being reabsorbed by the second qubit. In the case where the broken link is placed at the middle position between the two spins, one can see that the spin excitation propagates for an equal distance with the initial and final dispersion relation, reminiscent of a photon echo: dispersion effects have thus no influence in the transfer. We show in Fig.~\ref{Sbfig:timereversal}(d) a numerical simulation of the protocol where we could achieve a state-transfer fidelity of $\mathcal{F}=98\%$ with two system spins separated by $d=398a$, the small error being attributed to the tail of the wave-packet which remains trapped in the left part of the spin chain ($j<j_a$). For comparison, we obtain a fidelity $\mathcal{F}=67\%$ when we apply the state-transfer protocol presented in last section, for the same separation $d$ (and the cosine dispersion relation as here).

\subsection{A two qubit quantum gate mediated by spin-spin collisions}\label{Sbsub:gate}

The state-transfer protocol involves the propagation of a single excitation in the waveguide and is thus insensitive to the nature, photon or magnon, of the waveguide excitation. We now show an example where we make use of the hard-core nature of spin waves to realize a quantum gate between two qubits. The gate we have in mind is described by:
\begin{eqnarray}
|g\rangle_1 |g\rangle_2 &\to & |g\rangle_1|g\rangle_2\nonumber, \\
|g\rangle_1 |e\rangle_2 &\to &  e^{i\Phi}|e\rangle_1 |g\rangle_2 \nonumber,\\
|e\rangle_1 |g\rangle_2 &\to &  e^{-i\Phi}|g\rangle_1 |e\rangle_2\nonumber, \\
|e\rangle_1 |e\rangle_2 &\to &  -|e\rangle_1 |e\rangle_2 \label{Sbeq:gate},
\end{eqnarray}
which together with arbitrary single qubit gates, forms a universal set of gates~\cite{SbSchuch2003}.

\begin{figure}[t]
\includegraphics[width=0.5\textwidth]{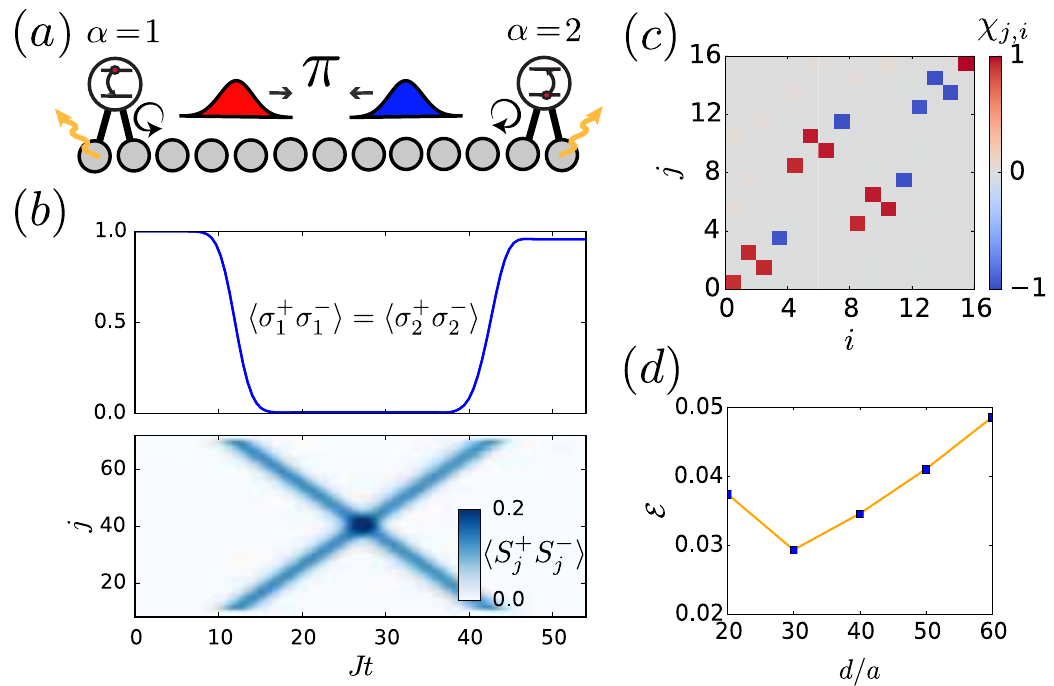}
\caption{Two qubit quantum gate via spin-spin collision. (a) Schematic representation of the entangling gate protocol. Two system spins with opposite chirality emit in opposite directions into the waveguide. Due to the spin nature of the waveguide, the magnonic wave-packets pick up a phase of $\pi$ in the collision, allowing for the realization of a phase gate, after the reabsorption of the excitations. (b) Numerical simulation of the gate for an initial state $e\rangle_1|e\rangle_2$ between two system spins, separated by $d=58a$, and the rest of the parameters as in Fig.~\ref{Sbfig:spintronics1} for the time-dependent couplings. The spin populations are represented in the upper panel. The lower panel shows the propagation of the two wave-packets in the spin waveguide. (c) Transfer matrix $\chi_{j,i}$ and (d) the error of the gate as a function of inter-spin distance $d$, showing the resilience of the protocol to dispersion effects.\label{Sbfig:gate}}
\end{figure}

The realization of such a gate in the context of our  spin model is shown schematically in Fig.~\ref{Sbfig:gate}(a). In the first part of the protocol, the two qubits are coupled to the waveguide via a time-dependent coupling of same magnitude $\tilde{J}_1(t)=\tilde{J}_2(t)$ and opposite phases $\phi_1=-\phi_2=\pi/4$, so that each qubit, if initially excited, emits a wave-packet towards the other qubit. As in Sec.~\ref{Sbsub:transfer}, the wave-packets are then absorbed using the coupling $\tilde{J}_1(\tau-t)=\tilde{J}_2(\tau-t)$ and the reversed phases $\phi_1=-\phi_2=-\pi/4$. Note that the time delay $\tau$ has to be sufficiently large so that the overlap between the emitting and absorbing pulses is negligible. 
If only one qubit is initially excited, the protocol reduces to the state-transfer presented in Sec.~\ref{Sbsub:transfer}, which also applies to photonic waveguides. However in the case where the two qubits are initially excited, the use of a spin waveguide becomes a crucial ingredient: the spin-chain state, described by a two excitation wave-function, acquires a $\pi$ phase sign when the two counter-propagating spin waves (emitted by both qubits) exchange positions~\cite{SbGorshkov2010}. Consequently, we obtain the required minus sign, which allows the entangling gate in Eqs.~\eqref{Sbeq:gate}.

We show in Fig.~\ref{Sbfig:gate}(b) a numerical simulation where the initial two-qubit state is $|e\rangle_1 |e\rangle_2$, corresponding to the case where the collision between the spin waves occurs. The two qubits exchange their populations by emitting simultaneously a wave-packet into the waveguide and reabsorbing the wave-packet coming from the other direction after the time-delay $\tau$. 

The gate efficiency can be assessed by calculating the transfer matrix $\chi$~\cite{SbPoyatos1997,SbChuang1997}, which relates any initial state $\rho_{\rm S}$ to the state $\rho'_{\rm S}$ obtained at the end of the protocol by $\rho'_j=\sum_{j,i}\chi_{j,i}\rho_i$. Here $(\rho_i)$ is defined by vectorizing $\rho_\mathrm{S}$ as $\rho_\mathrm{S}=\sum_i \rho_i A_i$, with the set $(A_i)=\lbrace|ee\rangle\langle ee|,|ee\rangle\langle eg|,..\rbrace$ being a basis of the (two-qubit) operator Hilbert space. We show in Fig.~\ref{Sbfig:gate}(c) the transfer matrix corresponding to the parameters in panel (b). The relative error $\mathcal{E}\equiv||\chi-\chi_\mathrm{P}||/||\chi_\mathrm{P}||$, with respect to the ideal transfer matrix $\chi_\mathrm{P}$ (calculated from Eq.~\eqref{Sbeq:gate}) is represented in Fig.~\ref{Sbfig:gate}(d) as a function of the distance $d$ between the qubits \footnote{Here, the trace norm is defined as for the BLP measure in Sec.~\ref{Sbsub:bandedge} as $||A||={\rm Tr}(\sqrt{A^\dag A})$.}. For these parameters, the trace distance between the two matrices remains very small, showing that the gate operates almost perfectly. As in the state-transfer protocol, the efficiency of the gate is, however, affected by dispersive effects at large distances. At very short distances, the gate can also not operate, as it would require to emit and absorb the two wave-packets simultaneously [cf.~Fig.~\ref{Sbfig:gate}(d)]. 

The present section has shown simulations of elementary quantum tasks in a quantum spin network, based on chiral coupling of qubits to the spin waveguide, involving control of the dispersion relation, and exploiting the natural interaction between the spin waves. In a broader context, this provides both the building blocks and the theoretical techniques to model complex composite quantum circuits based on spin waveguides as the communication channel. We will pursue this in future publications.

\section{Conclusions and Outlook}

In this article we have developed a theory of chiral quantum networks with spin chains as waveguides. The unique features of our model is an engineered chiral coupling of two-level systems, representing the nodes of the network to the spin waveguides representing the quantum channel. The main focus of our work has been the non-Markovian open system dynamics beyond the Born-Markov approximation, familiar from quantum optical descriptions. The physical origin of non-Markovianity in our model system is the nonlinear Bloch band dispersion relation for spin excitations, their finite propagation speed resulting in time delays in communication between the nodes, and/or strong coupling of the nodes to the spin waveguide. We have developed a description of the dynamics of such networks within an {\em extended Markovian model} (extended ME), where we keep the dynamics of the nodes and part of the spin chains connecting the nodes, while the `Markovian cut' is moved to the input and output ports of the network. The description of the waveguide by a spin chain offers a natural representation within tDMRG methods, which allows for efficient solution of the extended ME. In particular, this method opens perspectives to address computationally challenging regimes, such as the quantum feedback problem or networks with long delay times and highly populated waveguides \cite{SbPichler2015b,SbGrimsmo2015,SbTabak2015}. As an example of the non-Markovian network dynamics we have discussed the transient and steady state regime of system spins coupled to a chiral spin waveguide, including the driven-dissipative formation of quantum dimers. Furthermore, we have demonstrated how the chiral spin chains can be exploited for the design of quantum spintronic circuits for quantum information applications.

The present setting of chiral quantum networks with spins (and photons) has attractive physical implementations with various physical platforms. In a companion paper \cite{SbImplementation}, we give details for possible realizations of chiral quantum spin networks with Rydberg atoms, polar molecules and magnetic atoms, showing how chirality can be obtained via gauge fields naturally present in the dipole-dipole interactions. We moreover discuss a related implementation in trapped ion crystals, where phonon vibrations form a non-interacting bosonic chiral waveguide. In this companion paper, we also generalize the present model to {\em long-range} dipolar spin-spin couplings inherent to these setups. The theory developed in the present article thus furnishes both a setup as well as the computational tools for constructing `on-chip' quantum-communication networks within existing physical platforms and state-of-the-art technology.

\section*{Acknowledgments}
We thank M.~Baranov for useful discussions. Some time-dependent numerical solutions were obtained using the QuTiP toolbox~\cite{SbJohansson20131234}, and MPS simulations were performed using the ITensor library \cite{Sbitensor}. Work at Innsbruck is supported by the ERC-Synergy Grant UQUAM, the SFB FOQUS of the Austrian Science Fund, and the Army Research
Laboratory Center for Distributed Quantum Information via the project SciNet. B.V.~acknowledges the Marie Curie Initial Training Network COHERENCE for financial support. T.R.~was supported in part by BECAS CHILE.

\appendix

\section{Chiral master equation for infinite spin waveguide}\label{Sbapp:infiniteMasterChiral}

In this section, we derive the chiral Markovian ME (\ref{SbchiralMaster}) for the reduced density matrix of the system spins $\rho_{\rm S}$, assuming they are chirally coupled to an infinite spin waveguide. Under the three assumptions (i)-(iii) explained in Sec.~\ref{Sbsec:markov} of the main text, we can apply the standard Born-Markov procedure \cite{SbLehmberg:1970jj,Sbgardiner2015} to the closed network of system spins and infinite spin waveguide, described by Hamiltonians (\ref{SbHS}), (\ref{SbHB}) and (\ref{SbHSB}). As a result of this adiabatic elimination of the waveguide degrees of freedom, we formally obtain
\begin{align}
\dot{\rho}_{\rm S}=&-i[H_{\rm S},\rho_{\rm S}]\!+\!\sum_{\alpha,\beta}\left\{ Q_{\alpha-\beta}[\sigma^-_\beta\rho_{\rm S},\sigma_\alpha^+]+{\rm H.c.}\right\}.\label{SbMasterMarkDef}
\end{align}
Here $Q_{\alpha-\beta}\!=\!\lim_{s\to i\Delta}F_{\alpha-\beta}(s)$ are the Markovian reservoir-mediated couplings, expressed as a limiting case of the general system-waveguide coupling functions 
\begin{align}
F_{\alpha-\beta}(s)=\!\int_{0}^{\pi/a}\!\!\!dk \,\frac{[g_k^2e^{ikd(\alpha-\beta)}+g_{-k}^2e^{-ikd(\alpha-\beta)}]}{s+i\omega_k}.
\end{align}
Quite remarkably, these coupling functions $F_{\alpha-\beta}(s)$ can be evaluated exactly in the present model, taking into account all reservoir modes and the non-linearities of the dispersion $\omega_k$ and coupling $g_k$, given in Sec.~\ref{Sbchirality_infinite}. In fact, using the integrals,
\begin{align}
I_1&=\int_{-1}^{1}\frac{dx}{\pi}\,\frac{(1+px)\cos(n\arccos(x))}{\sqrt{1-x^2}(z-ix)}\nonumber\\
&=\frac{(1-ipz)}{\sqrt{1+z^2}}e^{in\arccos(-iz)}+ip\delta_{n0},\\
I_2&=\int_{-1}^{1}\frac{dx}{\pi}\,\frac{\sin(n\arccos(x))}{(z-ix)}=-ie^{in \arccos(-iz)},
\end{align}
valid for ${\rm Re}(z)>0$ and integer $n\geq 0$, the general couplings $F_{\alpha-\beta}(s)$, for ${\rm Re}(s)>0$, take the form
\begin{align}
F_{\alpha-\beta}(s)\!={}\!\frac{\tilde{J}^2}{J}\!\!&\left[\frac{1\!-\!i\cos(2\phi)\bar{s}}{\sqrt{1+\bar{s}^2}}\!+\!(2\bar{\Theta}_{\alpha-\beta}\!-\!1)\sin(2\phi)\!\right.\nonumber\\
&\left.+i\delta_{\alpha\beta}\cos(2\phi)\right]e^{i|\alpha-\beta|(d/a)\arccos(-i\bar{s})}.\label{SbgeneralCouplings}
\end{align}
Here $\bar{s}=s/2J$, $\bar{\Theta}_{\alpha-\beta}$ is the Heaviside function defined such that $\bar{\Theta}_{0}=1/2$ and $\delta_{\alpha\beta}$ is the Kronecker delta.

In Appendix~\ref{Sbapp:WW}, the $s$-dependence of $F_{\alpha-\beta}(s)$ will be crucial to describe non-Markovian effects in the interaction of system spins with the waveguide. However, in the Markovian regime, the couplings $Q_{\alpha-\beta}$ are obtained by taking the limit $s\rightarrow i\Delta$ in Eq.\,(\ref{SbgeneralCouplings}), which yields
\begin{align}
Q_{\alpha-\beta}\!&=[\gamma_R\bar{\Theta}_{\alpha-\beta}+\gamma_L\bar{\Theta}_{\beta-\alpha}]e^{i\bar{k}d|\alpha-\beta|}-i\omega_{\rm LS}\delta_{\alpha\beta}.\label{SbmarkCouplings}
\end{align}
Here, $\gamma_L$ and $\gamma_R$ are the asymmetric decay rates into left and right moving reservoir modes, given in Eq.~(\ref{SbdecayGeneral}) of the main text (for $J>0$). In addition, the Lamb shift
\begin{align}
\omega_{\rm LS}={\rm P}\int_{-\pi/a}^{\pi/a}dk \frac{|g_k|^2}{\omega_k+\Delta_{\rm S}}=-\frac{\tilde{J}^2}{J}\cos(2\phi),
\end{align}
renormalizes the transition frequency of the system spins as $\Delta\rightarrow\tilde{\Delta}=\Delta+\omega_{\rm LS}$, and $\bar{k}a=\arccos(\tilde{\Delta}/2J)$ is the resonant wavevector of the right moving excitations, satisfying $v_{\bar{k}}>0$. Finally, replacing Eq.\,(\ref{SbmarkCouplings}) into Eq.\,(\ref{SbMasterMarkDef}) and rearranging terms, we obtain the chiral ME in Lindblad form as shown in Eq.\,(\ref{SbchiralMaster}), with reservoir-mediated coherent interactions and collective jump operators as given in Eqs.~(\ref{SbchiralH}) and (\ref{SbglobalJump}), respectively.

Finally, we notice that in the case of inhomogeneous detunings of the system spins, $\Delta_\alpha=\Delta+\delta\Delta_\alpha$, the previous derivation of the ME still applies provided these inhomogeneities are smaller than the bandwidth, i.e.~$|\delta\Delta_\alpha|\lesssim\gamma_R+\gamma_L$. This is particularly important for the formation of multi-partite entangled clusterized phases of the system spins in steady state, as discussed in Refs.~\cite{SbStannigel:2012jk,SbPichler2015}.

\section{Chiral master equation for a finite spin waveguide with losses: Proof for perfect absorbing boundaries in the weak coupling limit}\label{Sbapp:FiniteMaster}

The goal of this subsection is to show that, in the presence of a finite waveguide with local losses at its ends, one can derive a ME for the system spins, which coincides with the corresponding one for an infinite chain, given in Eq.~\eqref{SbchiralMaster} of the main text. To do so, we first divide the total Hilbert space in the set of subspaces $P^{(i)}$, containing exactly $i$ bath excitations. In the Born-Markov approximation, we can neglect the excitation of the manifolds $P^{(i>2)}$ and adiabatically eliminate $P^{(1)}$ to obtain a ME describing the effective dynamics of the system spins in the slow subspace $P^{(0)}$. Applying the formalism developed in Ref.~\cite{SbReiter2012} to the extended ME (\ref{SbfullMaster}), we obtain
\begin{equation}
\dot{\rho}_\mathrm{S}=-i[H_\mathrm{S}+H_\mathrm{C},\rho_{S}]+\sum_{X=L,R}\sum_{n=0}^{M-1}\mathcal{D}[\mathcal{L}_{X(n)}]\rho_\mathrm{S}\label{Sbeq:master_elim}
\end{equation}
where the effective Hamiltonian and jump operators read $H_\mathrm{C}=-\frac{1}{2}V_{-}\mathcal{R}V_{+}$ and $\mathcal{L}_{X(n)}=\sqrt{\Gamma_n}S^-_{X(n)}H_{\mathrm{NH}}^{-1}V_{+}$, with $L(n)=1+n$, and $R(n)=N_\mathrm{B}-n$. The operator $V_{-}=P^{(0)}H_\mathrm{SB}P^{(1)}$ transfers an excitation from the spin chain to the system ($V_+=V_-^\dagger$) and the non-Hermitian bath Hamiltonian is defined as
\begin{align}
H_{\mathrm{NH}}=P^{(1)}(H_\mathrm{B}-\sum_{n,X}\tfrac{i\Gamma_n}{2}S_{X(n)}^+S^-_{X(n)})P^{(1)}.
\end{align}
The other auxiliary quantities read $\mathcal{R}=H_{\mathrm{NH}}^{-1}+\left(H_{\mathrm{NH}}^{-1}\right)^\dagger$. Using Eq.~\eqref{SbHB}, we can write $H_\mathrm{NH}$ as a $N_\mathrm{B}\times N_\mathrm{B}$ matrix 
\begin{equation}
(H_\mathrm{NH})_{j,l}=-J\delta_{j,l\pm1}-\sum_n\frac{i\Gamma_n}{2}\delta_{j,l}(\delta_{j,1+n}+\delta_{j,N_\mathrm{B}-n}),\label{Sbnonhermitionexplicit}
\end{equation}
while the coupling $V_-$ can be written in the form of a $N_\mathrm{S}\times N_\mathrm{B}$ matrix as $(V_-)_{\alpha,j}=\tilde{J}\left(e^{-i\phi}\delta_{j,R[\alpha]}+e^{i\phi}\delta_{j,L[\alpha]}\right)$. In the general case, the inverse $H_{\rm NH}^{-1}$ must be numerically calculated in order to obtain the system ME via Eqs.~\eqref{Sbeq:master_elim}. However, in the special case of a single loss per end of the chain, $M=1$, and an even number of bath spins $N_\mathrm{B}$, the non-Hermitian Hamiltonian (\ref{Sbnonhermitionexplicit}) can be expressed as the sum of a symmetric tridiagonal Toeplitz (STT) matrix~\cite{SbFischer1969} and a rank 2 matrix, whose inverse can be analytically calculated using the Woodbury identity. Finally, for $\Gamma_0=2J$, we obtain a particularly simple result of the inverse
\begin{equation}
(H^{-1}_\mathrm{NH})_{j,l} = \frac{i}{2 J}e^{i \pi  |j-l|/2}.
\label{Sbeq:invtwosinks2J}
\end{equation}
Plugging Eq.~\eqref{Sbeq:invtwosinks2J} in Eq.~\eqref{Sbeq:master_elim}, we exactly obtain the chiral ME \eqref{SbchiralMaster}, derived in Appendix~\ref{Sbapp:infiniteMasterChiral} assuming an infinite spin waveguide. This identification proves that in the Markovian approximation, a finite spin chain with a single loss of $\Gamma_0=2J$ at each of its ends, can simulate perfect absorbing boundary conditions \cite{SbGivoli:1991js} and thus the physics of an infinite reservoir. Although we have only derived this result for an even number of bath spins $N_{\rm B}$, we checked using symbolic numerical libraries~\cite{SbSymPyDevelopmentTeam2014} that it also applies for the odd case. 

Finally, we note that the waveguide observables can also be derived from the adiabatic elimination technique. For example, the bath spin amplitude at site $j$ is simply given by
\begin{equation}
\langle S_j^- \rangle = \sum_\alpha (H^{-1}_\mathrm{NH}V_+)_{j,\alpha}\langle \sigma_\alpha^-\rangle.
\end{equation}

\section{Wigner-Weisskopf treatment of undriven system spins coupled to a chiral spin waveguide}\label{Sbapp:WW}

In this section, we consider the spin network model in the absence of driving ($\Omega_\alpha=0$), and find analytical solutions for the dynamics of the system spins beyond the Markovian approximation. For an initial condition in the global $1$-excitation manifold, the state $|\Psi(t)\rangle$ for system spins and waveguide can be exactly written at all times in a Wigner-Weisskopf ansatz as 
\begin{align}
|\Psi(t)\rangle=\left(\sum_\alpha c_{\alpha}(t)\sigma_\alpha^++\int\!dk\ c_{k}(t)b_k^\dag\right)|g\rangle^{\otimes N_{\rm S}}|0\rangle.\label{SbnoDriveAnsatz}
\end{align}
Here, $c_\alpha(t)$ and $c_k(t)$ are the probability amplitudes describing the presence of the excitation in the system spin $\alpha$ and in the spin wave momentum state $k$, respectively. Replacing Eq.\,(\ref{SbnoDriveAnsatz}) into the Schroedinger equation $d|\Psi(t)\rangle/dt=-i(H_{\rm S}+H_{\rm B}+H_{\rm SB})|\Psi(t)\rangle$, with the Hamiltonians given in Sec.~\ref{Sbsec:model}, one obtains the following coupled differential equations for the amplitudes
\begin{align}
\dot{c}_\alpha(t)&=i\Delta c_\alpha(t)-i\int\!\!dk\,e^{i\alpha kd}c_k(t),\\
\dot{c}_k(t)&=-i\omega_k c_k(t)-ig_k\sum_\alpha e^{-i\alpha kd}c_\alpha(t).\label{Sbphotonfield}
\end{align}
Writing the latter equations in Laplace space and solving for $\tilde{c}_\alpha(s)=\int_0^\infty dt\,e^{-st} c_\alpha(t)$, one obtains
\begin{align}
&\left[s\!-\!i\Delta\!+\!F_0(s)\right]\!\tilde{c}_\alpha(s)+\!\sum_{\beta\neq \alpha}\!F_{\alpha-\beta}(s)\tilde{c}_\beta(s)=c_\alpha(0),\label{SbsystemEqnoDrive}
\end{align}
where the functions $F_{\alpha-\beta}(s)$ are given in Eq.~(\ref{SbgeneralCouplings}), and we assumed that the bath spins are initially not excited, $c_k(0)=0$.

The linear system of equations (\ref{SbsystemEqnoDrive}) describes the dynamics of $N_{\rm S}$ system spins, with interactions mediated by the infinite spin waveguide in the $1$-excitation manifold. For simplicity, we consider the case $N_{\rm S}\leq 2$, whose solution reads
\begin{align}
\tilde{c}_1(s)&=\frac{1}{[1-B(s)]}\!\left[\frac{c_1(0)}{s-i\Delta\!+\!F_0(s)}\!-\!\frac{c_2(0)B(s)}{F_1(s)}\right],\label{Sbc1noDrive}\\
\tilde{c}_2(s)&=\frac{1}{[1-B(s)]}\!\left[\frac{c_2(0)}{s-i\Delta\!+\!F_0(s)}\!-\!\frac{c_1(0)B(s)}{F_{-1}(s)}\right],\label{Sbc2noDrive}
\end{align}
where
\begin{align}
B(s)=\frac{F_1(s)F_{-1}(s)}{[s-i\Delta+F_0(s)]^2}.
\end{align}
In the following Secs.~\ref{Sbsubsub:bandedge} and \ref{Sbsubsub:retardation}, we analytically perform the inverse Laplace transform of the functions (\ref{Sbc1noDrive})-(\ref{Sbc2noDrive}), via the integral
\begin{align}
c_\alpha(t)=\frac{1}{2\pi i}\lim_{y\to\infty}\int_{\epsilon-iy}^{\epsilon+iy}\tilde{c}_\alpha(s)e^{st}ds,\qquad (\epsilon>0),\label{SbinverseLaplace}
\end{align}
and thus solve for the corresponding non-Markovian dynamics in the case of one and two emitters, respectively.

\subsection{Strong coupling in the single emitter problem: Band-edge physics and chirality}\label{Sbsubsub:bandedge}

\begin{figure}[t]
\includegraphics[width=\columnwidth]{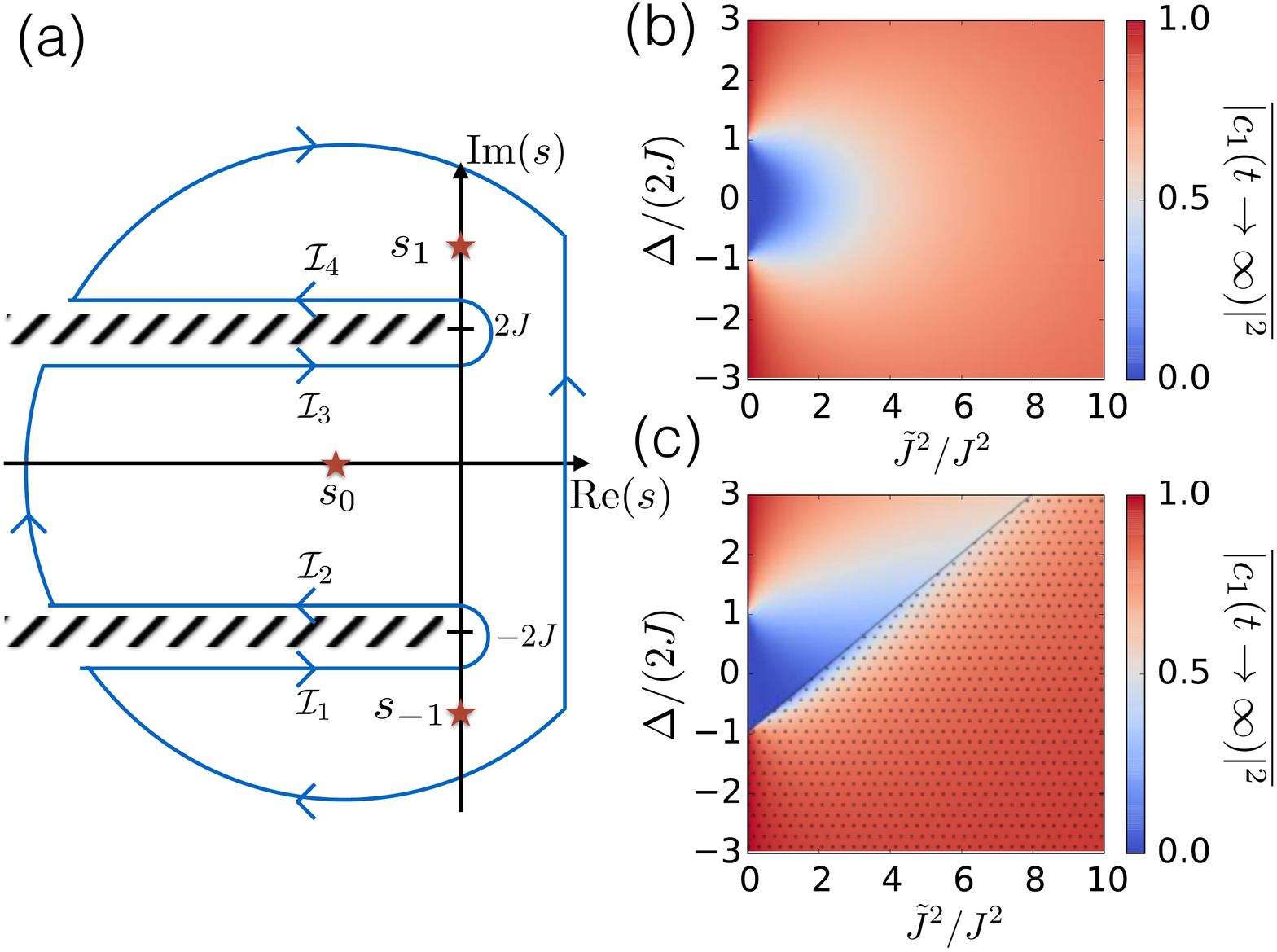}
\caption{General solution of a single emitter chirally coupled to the spin waveguide. (a) Contour integral used in the calculation of $c_1(t)$ [cf.~Eq.~\eqref{Sbeq:ctpole}]. (b)-(c) Phase diagram of the chiral bound state problem as a function of $(\tilde{J}/J)^2$ and $\Delta$, for $\phi=\pi/4$ (b) and $\phi=0$ (c).}\label{Sbfig:contour}
\end{figure}

In this section we analytically solve the problem of a single emitter coupled to the chiral waveguide, which in the strong coupling regime ($\tilde{J}\gtrsim J$) exhibits band-edge effects associated to the formation of localized bound states between system and waveguide \cite{SbJohn1990,SbKofman1994,SbDouglas2015,SbCalajo2015}. Assuming $N_{\rm S}=1$ in Eq.\,(\ref{SbgeneralCouplings}), we obtain $F_{\pm 1}(s)=B(s)=0$, and thus the problem in Eqs.~(\ref{Sbc1noDrive})-(\ref{Sbc2noDrive}) with $c_1(0)=1$, reduces to Laplace inverting the function $\tilde{c}_1(s)=1/[s-i\Delta+F_0(s)]$, where
\begin{align}
F_0(s)&=\frac{\tilde{J}^2}{J}\left(\frac{[1-i\cos(2\phi)\bar s]}{\sqrt{1+{\bar s}^2}}+i\cos(2\phi)\right).
\end{align}
The calculation of the inverse Laplace transform requires to define the branch cuts of the multivalued function $\sqrt{1+\bar{s}^2}$ in the complex plane. Here, we choose 
\begin{equation}
\sqrt{1+\bar s^2}\equiv\sqrt{|1+\bar s^2|}\exp\left[\frac{\mathrm{arg}(\bar s+i)+\mathrm{arg}(\bar s-i)}{2}\right],
\end{equation}
corresponding to two branch cuts at $\bar s=s/2J=\pm i$. Accordingly, we choose the contour integral defined in Fig.~\ref{Sbfig:contour}(a) and obtain via the Residue theorem
\begin{equation}
c_1(t)=\sum_{n=-1,0,1}\mathrm{Res}[\tilde{c}_1,s_n]e^{s_nt}+\frac{1}{2\pi i}\sum_{m=1}^4 \mathcal{I}_m.\label{Sbeq:ctpole}
\end{equation}
Here, $s_n$ denote the poles of $\tilde{c}_1(s)$ satisfying
\begin{equation}
s-i\Delta+F_0(s)=0\label{Sbeq:pole},
\end{equation}
and $\mathcal{I}_m$ label the various integrals of $\tilde{c}_1(s)e^{st}$ along the four branch cuts shown in Fig.~\ref{Sbfig:contour}(a). Finally, the residues associated to each pole $s_n$ are given by
\begin{equation}
\mathrm{Res}[\tilde{c}_1,s_n]=\frac{1}{1+F'_0(s_n)}.\label{Sbeq:residue}
\end{equation}

Notice that the pole decomposition in Eq.~(\ref{Sbeq:ctpole}) allows us to understand the behavior of the system dynamics in simple terms: a real pole corresponds to the usual exponential decay behavior, whereas the existence of purely imaginary poles is associated to the presence of (non-decaying) bound states (as also shown in Fig.~\ref{Sbfig:NonMark_time_evo}). Finally, the contributions of the branch cuts ${\cal I}_m$ vanish at long-time $t\to\infty$, showing an oscillating behavior in the transient.

In the case of perfect chirality $\phi=\pi/4$, and a resonant coupling $\Delta=0$, the poles and residues can be calculated analytically. This allows us to derive the expression for $c_1(t)$ in Eq.~(\ref{Sbeq:boundstate}) of the main text, valid for arbitrary strong coupling $\tilde{J}/J$. Specifically, the real pole leading to a modified decay rate is given by $s_0=-\sqrt{2(\lambda-1)}J$ with $\lambda=[1+(\tilde{J}/J)^2]^{1/2}$, whereas the two imaginary poles associated to the bound state excitation read $s_{\pm 1}=\pm i \sqrt{2(\lambda+1)}J$. In addition, the corresponding residues are given by $\mathrm{Res}[\tilde{c}_1,s_0]=(\lambda+1)/(2\lambda)$, and $\mathrm{Res}[\tilde{c}_1,s_{\pm 1}]=(\lambda-1)/(2\lambda)$. The integrals along the branch cut are non-analytic \cite{SbTanaka:2006cn,SbNavarrete-Benlloch2011}, however we can characterize their long-time behavior considering the limit $\mathrm{Re}[s]\to0$. For instance, using $\int_{0}^{\infty}e^{xt}\sqrt{x}=\sqrt{\pi}/(2 t^{3/2})$ and $\bar t=2J t$, we obtain $\mathcal{I}_3+\mathcal{I}_4\approx -\sqrt{\pi}J^2/\tilde{J}^2(2i/\bar t)^{3/2} e^{i\bar t}$, for ${t\to\infty}$, showing that this contribution decays with a power-law behavior $\sim \bar{t}^{-3/2}$ and oscillates with the band-edge frequency $2J$. When neglecting these contributions for $Jt\gg 1$, we obtain the expression in Eq.~(\ref{Sbeq:boundstate}) of the main text.

In the general chiral case with $\phi\neq \pi/4$ and $\Delta\neq 0$, we solve the pole equation Eq.~\eqref{Sbeq:pole} numerically and obtain the corresponding residues from  Eq.~\eqref{Sbeq:residue}. The purely imaginary poles, associated to bound state solutions, determine the behavior of the system spin dynamics in the very long time limit as
\begin{equation}
c_1(t\to\infty)=\sum_n \mathrm{Res}[\tilde{c}_1,s_n] e^{s_n t},
\end{equation}
where the sum is taken over purely imaginary poles. In the case where only one bound state is present, the steady state behavior is characterized by a constant two-level atom population $|c_1|^2=|\mathrm{Res}[\tilde c_1,s_n]|^2$ representing the overlap between the initial state and the bound state solution. In the case where two or more bound states are present, the system spin population $|c_1|^2$ oscillates between $0$ and $(\sum_n|\mathrm{Res}[\tilde c_1,s_n]|)^2$.
 
The average system spin occupation in steady state $\overline{|c_1(t\to \infty)|^2}=\sum_n |\mathrm{Res}[\tilde c_1,s_n]|^2$ is shown in Figs.~\ref{Sbfig:contour}(b) and (c) for the unidirectional ($\phi=\pi/4$), and bidirectional ($\phi=0$) cases, respectively, as a function of the detuning $\Delta$ and the relative coupling strength $\tilde{J}^2/J^2$. The blue region identifies the Markovian limit where the spin behavior is characterized by a decay into the waveguide whereas the red region corresponds to the parameters for which the bound state excitation is dominant. In the cascaded case [cf.~Fig.~\ref{Sbfig:contour}(b)], we find that the steady state is always characterized by a superposition of two bound states whereas in the bidirectional limit [cf.~Fig.~\ref{Sbfig:contour}(c)], we identify a region with only one bound state (shown by black dots).

\subsection{Retardation effects for two emitters chirally coupled to a waveguide}\label{Sbsubsub:retardation}

In this section we study the case of two undriven system spins chirally coupled to a waveguide, and separated by a distance $d$, such that they interact via emitting and absorbing waveguide excitations with a time delay $\tau$. The corresponding dynamics, in the absence of driving $\Omega_\alpha=0$ and for a single global excitation, is obtained by Laplace inverting the functions in Eqs.~(\ref{Sbc1noDrive})-(\ref{Sbc2noDrive}). In the bidirectional case ($\gamma_{\rm R}=\gamma_{\rm L}$), this problem was solved by Milonni and Knight in Ref.~\cite{SbMiloni1974}, but here we generalize the solution to the chiral case $\gamma_{\rm R}\geq\gamma_{\rm L}$. Following the original approach, we expand in geometric series $[1-B(s)]^{-1}=\sum_{n=0}^\infty B^n(s)$ in Eqs.~(\ref{Sbc1noDrive})-(\ref{Sbc2noDrive}) (provided $|B(s)|<1$), and thus obtain a convenient expression for $\tilde{c}_\alpha(s)$, with $\alpha=1,2$, as 
\begin{align}
\tilde{c}_\alpha(s)&=\delta_{\alpha2}c_{\alpha}(0)\tilde{R}_{011}(s)\!+\!\sum_{\beta=1}^{2}c_\beta(0)\!\sum_{n=0}^{\infty}\!\tilde{R}_{n\alpha\beta}(s).\label{SbcalphaNoDrive}
\end{align}
Here, the Laplace functions $\tilde{R}_{n\alpha\beta}(s)$, are explicitly given by
\begin{align}
\tilde{R}_{n\alpha\beta}(s)=(-1)^{\alpha+\beta}\frac{[F_1(s)]^{n+\alpha-1}[F_{-1}(s)]^{n+\beta-1}}{[s-i\Delta+F_0(s)]^{2n+\alpha+\beta-1}}.
\end{align}
As $\tilde{c}_\alpha(s)$ is now written as a sum over $\tilde{R}_{n\alpha\beta}(s)$ functions, one can clearly see that the corresponding poles are identical to the single emitter case $N_{\rm S}=1$, satisfying Eq.~(\ref{Sbeq:pole}), but here they appear in arbitrary higher orders. Analytically calculating the corresponding higher order residues is very challenging, but in the weak coupling limit $\tilde{J}/J\ll 1$, we can neglect the contribution from the imaginary poles and expand $\tilde{R}_{n\alpha\beta}(s)$ in powers of $|s|/2J\sim (\tilde{J}/J)^2\sim \Delta/2J\sim (d/a)(\tilde{J}/J)^2\ll 1$. To first order, we obtain a simple expression that admits an analytical Laplace inverse:
\begin{align}
\tilde{R}_{n\alpha\beta}(s)\approx (-1)^{\alpha+\beta}\frac{\gamma_{\rm R}^{n+\alpha-1}\gamma_{\rm L}^{n+\beta-1}e^{i\pi J\tau_{n\alpha\beta}}e^{-s\tau_{n\alpha\beta}}}{[s+\gamma/2-i\tilde{\Delta}]^{2n+\alpha+\beta-1}},\label{SbsimpleR}
\end{align}
with $\tau_{n\alpha\beta}=(2n+\alpha+\beta-2)\tau$ a generalized retarded time. Notice that Eq.~(\ref{SbsimpleR}) neglects the coupling to band-edge modes, and simply corresponds to the case of a 1D photonic bath with linear dispersion relation, as in Ref.~\cite{SbMiloni1974}. By Laplace inverting $\tilde{R}_{n\alpha\beta}$, we obtain the general solution for the dynamics of both system spins $\alpha=1,2$, including retardation as
\begin{align}
c_\alpha(t)&=\delta_{\alpha2}c_{\alpha}(0)R_{011}(t)+\sum_{\beta=1}^{2}c_\beta(0)\!\sum_{n=0}^{\infty}\!R_{n\alpha\beta}(t-\tau_{n\alpha\beta}),\label{SbSolutioncalphaNoDrive}
\end{align}
where
\begin{align}
R_{n\alpha\beta}(t)={}&(-1)^{\alpha+\beta}\frac{\gamma_{\rm R}^{n+\alpha-1}\gamma_{\rm L}^{n+\beta-1}e^{i\bar{k}d(2n+\alpha+\beta-2)}}{(2n+\alpha+\beta-2)!}\nonumber\\
&\times e^{-(\gamma/2+i\tilde{\Delta})t}t^{2n+\alpha+\beta-2}\Theta(t).\label{SbTnalphabeta}
\end{align}
Finally, the expressions (\ref{SbWWret1})-(\ref{SbWWret2}) in the main text are obtained as a specific case of Eq.~(\ref{SbSolutioncalphaNoDrive}) by assuming $\tilde{\Delta}=0$ and the initial condition $c_1(0)=1$ and $c_2(0)=1$, in addition to the identifications $f_n^{(m)}(t)=R_{n,m+1,1}(t)$, $\tau_{n,m+1,1}=(2n+m)\tau$, and $\tau=d/(2Ja)$. 

\section{Details on the MPS calculations with the quantum trajectories approach}\label{Sbapp:MPS_entropy}

In this appendix, we give more details on the MPS approach used to calculate the dynamics shown in Fig.~\ref{Sbfig:longdelay} and analyze in particular the role of the bosonic/spin character of the waveguide.

\begin{figure}[t]
\includegraphics[width=\columnwidth]{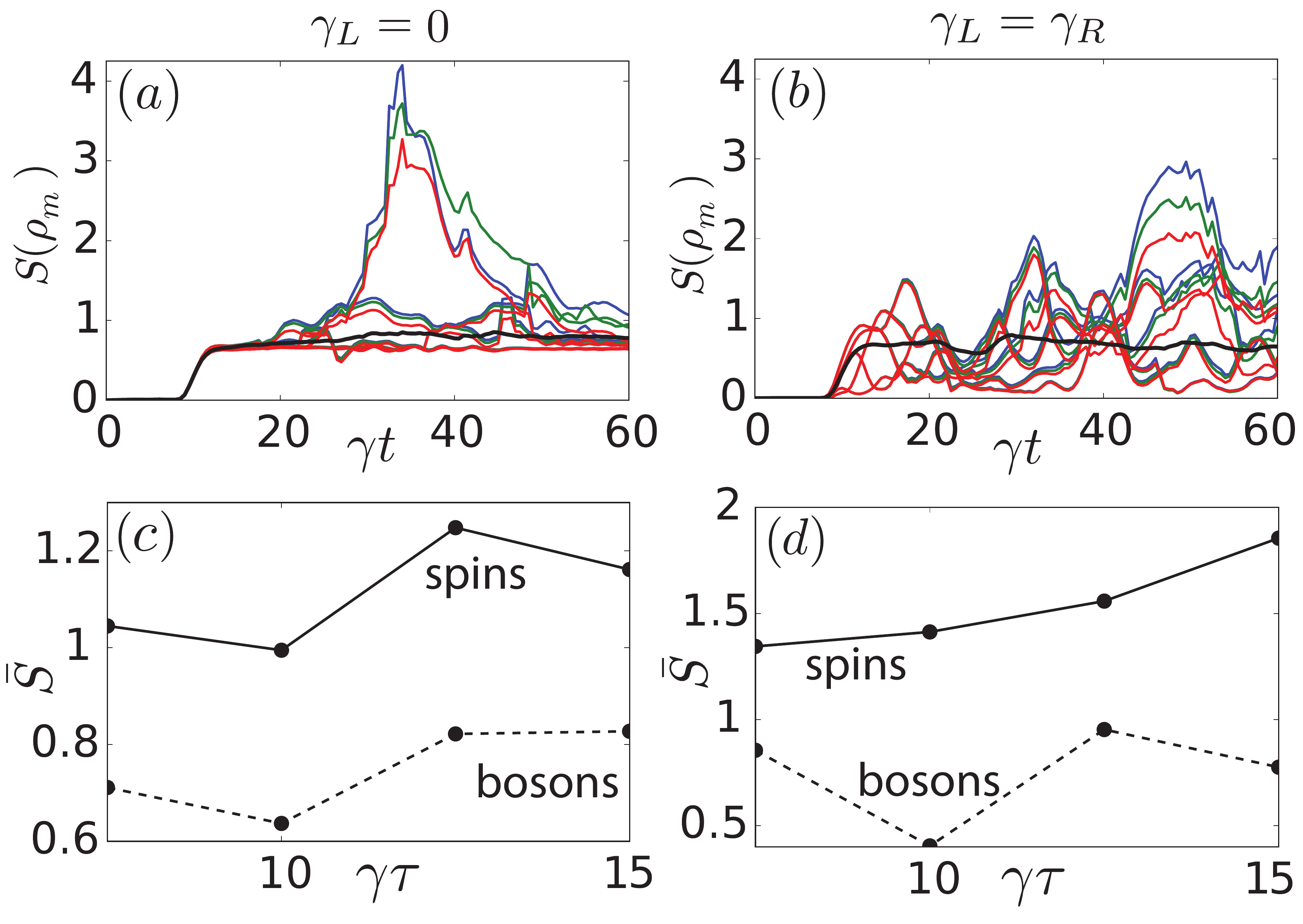}
\caption{(a,b) Entropies $S(\rho_m)$ for a partition in the middle of a bosonic waveguide, for a representative sample of MPS quantum trajectories, and calculated for three different bond dimensions $D=30,60,120$ (shown as red, green, and blue lines, respectively). Parameters are the same as in Figs.~\ref{Sbfig:entropytrajectories}(a,b), except for the bosonic nature of the waveguide $S_j^-\to b_j$. The average entropy $\bar{S}(t)$ is represented by a black solid line. For the unidirectional bosonic waveguide (a), the behavior is analogous to the spin case in Fig.~\ref{Sbfig:entropytrajectories}(a) with the majority of the trajectories staying at low entropy, but with some of them reaching high values at long times. In the bidirectional case (b), the trajectories have a larger average entropy compared to (a), but a much smaller one with respect to its spin counterpart in Fig.~\ref{Sbfig:entropytrajectories}(b). (c,d) Comparison of the average entropy $\bar{S}(2\tau)$ for spin and boson waveguides, and as a function of $\gamma\tau$. We assume the same situation and parameters as in Fig.~\ref{Sbfig:longdelay}, with $D=30$ and $\tilde{J}=0.5J$, but we vary $\gamma\tau$ by choosing the distance between system spins as $d/a=30,40,50,60$. The average entropy $\bar{S}(2\tau)$ increases with $\tau$ (up to an oscillation with $\bar{k}d$), and is larger for a spin than for a boson waveguide due to the hard-core constraint.}
\label{Sbfig:boson_average}
\end{figure}

Analogously to Fig.~\ref{Sbfig:entropytrajectories}, but for a bosonic waveguide, we show in Figs.~\ref{Sbfig:boson_average}(a,b) the entropies $S(\rho_m)$ as a function of time, for a representative sample of trajectories, and three different maximum bond dimensions $D=30,60,120$. In the case of unidirectional coupling, the trajectories are in general associated with a low level of entropy as in the spin case. For a bidirectional boson waveguide, the average entropy is larger than the unidirectional case, but it is much smaller than for its spin counterpart in Fig.~\ref{Sbfig:entropytrajectories}(b). In both cases, the presence of a few trajectories with very large entropy limits our method at long times, as they cannot be correctly described with a small bond dimension. 

Additionally, we show in Figs.~\ref{Sbfig:boson_average}(c,d) the average entropy $\bar{S}(2\tau)$ as a function the time delay $\tau\propto d$, and compare the results for a spin (solid) and a bosonic (dashed) waveguide. For a unidirectional coupling, the entropy stays on the order of $1$, with an oscillating behavior, which we attribute to the different phases $\bar{k}d$ accumulated by the waveguide excitations during the propagation between the two system spins. In the bidirectional case, the entropy increases with the time delay $\tau$ as the existence of two channels gives the possibility to entangle the system spins with waveguide excitations propagating in both directions~[cf.~Ref.~\cite{SbPichler2015b} for a study in the photonic context]. For a spin waveguide, in particular, the average entropy can reach large values as each collision between spin waves can be associated with an interaction phase shift of $\pi$ (hard-core constraint). The number of these events increase with the delay $\tau$, which is associated to the entropy growth observed in Fig.~\ref{Sbfig:boson_average}(d).

With this particular example, we have shown that the quantum trajectories MPS approach can simulate the dynamics of our chiral network over long times and with small bond dimensions. The losses placed at the ends of the waveguide play in this context an important role as they dissipate the entropy which is irrelevant for the system spin dynamics (excitations leaving the waveguide, for instance). A limitation of the present MPS method is the large bond dimension required by some trajectories at long times, and thus it would be interesting to test other MPS approaches on our network model, specially the ones tailored to perform the evolution directly in the density matrix representation~\cite{SbWerner:2014wq,SbCui:2015im,SbMascarenhas:2015cc}.

\section{Exact linear dispersion relation via long-range interactions between bath spins}\label{Sbapp:triangulardispersion}

We present in this appendix the bath spin couplings, which are required to obtain an exactly linear dispersion relation for the magnons in the waveguide. In particular, a triangular dispersion relation of the form $\omega(k)=2J|k|$, can be obtained by simply adding long-range interactions in the bath Hamiltonian as
\begin{equation}
H_\mathrm{B}=-\frac{4J}{\pi}\sum_{j,n} (2n+1)^{-2}S^+_jS_{j+2n+1}^-+\mathrm{H.c}.
\end{equation}
Then, by expressing this Hamiltonian in terms of momentum eigenstates $b_k$ as in Eq.~(\ref{SbHBbulk}), we obtain the expected dispersion relation. 

\section{State transfer pulses}\label{Sbapp:gaussianpulse}

Here we present the time-dependent couplings used to generate the Gaussian wave-packets in Sec.~\ref{Sbsec:spintronics}. Our method is based on Ref.~\cite{SbStannigel2011}, where the shape of the required couplings are derived analytically under the weak-coupling assumption, so that the dispersion relation can be approximated as linear. Specifically, we choose $\tilde{J}_1(t)=\sqrt{\gamma_1(t)J/2}$ and $\tilde{J}_2(t)=\tilde{J}_1(\tau-t)$, with $\gamma_1(t)=(2\sqrt{\kappa}\text{\ensuremath{\gamma}}_{m}e^{-\kappa(t-t_{m})^{2}})/(2\sqrt{\kappa}-\sqrt{\pi}\text{\ensuremath{\gamma}}_{m}\text{erf}(\sqrt{\kappa}(t-t_{m})))$. Further assuming $\kappa=(1.01\pi\gamma_m^2)/4$ and $t_m=6/\gamma_m$, one obtains an ideal maximum transfer fidelity $\mathcal{F}_{\rm ideal}\equiv\lim_{t\to\infty}\langle \sigma^+_2\sigma_2^-(t)\rangle \sim0.995$, limited only by the finite duration of pulse. In the examples presented in this work we considered $\gamma_m=0.3J$.

\bibliographystyle{apsrev4-1}
%

\end{document}